\documentclass{elsarticle}
\usepackage{amsmath,amsthm,amssymb}
\usepackage{graphicx,tikz,epsfig,caption,musixtex,minibox,multirow,makecell}
\usepackage{pgfplots}
\usepackage{rotating}
\usepackage{relsize}
\tikzset{fontscale/.style = {font=\relsize{#1}}
    }
\usepackage[shortlabels]{enumitem}
\DeclareMathOperator*{\argmin}{arg\,min}

\usepackage{pdfpages}
\usepackage{blkarray}
\usetikzlibrary{decorations.markings}
\usetikzlibrary{calc}

\usepackage[utf8]{inputenc}

 \usepackage{graphics}
 \usepackage{tikz}
 \usetikzlibrary{arrows}
  \usepackage{listings}

\usepackage{amssymb}

\usepackage{cancel}

\usepackage{algorithm}
\usepackage[noend]{algcompatible}
\usepackage[noend]{algpseudocode}
\usepackage{algpseudocode}

\usepackage{kotex}

\newcommand{\figref}[1]{Figure~\ref{#1}}
\newcommand{\tabref}[1]{Table~\ref{#1}}

\theoremstyle{plain}
\newtheorem{theorem}{Theorem}[section]

\newtheorem{proposition}[theorem]{Proposition}

\theoremstyle{definition}
\newtheorem{definition}[theorem]{Definition}

\theoremstyle{remark}
\newtheorem{remark}[theorem]{Remark}
\newtheorem{example}[theorem]{Example}

\begin{document}

\begin{frontmatter}

\title{Machine Composition of Korean Music via Topological Data Analysis and Artificial Neural Network}
\author{Mai Lan Tran \textsuperscript{ab}, Dongjin Lee\textsuperscript{cb}, Jae-Hun Jung \textsuperscript{ab}$^{\ast}$ \\
{\textsuperscript{a} Department of Mathematics, Pohang University of Science \& Technology, Pohang 37673, Korea; \\
\textsuperscript{b} POSTECH Mathematical Institute for Data Science (MINDS), Pohang University of Science \& Technology, Pohang 37673, Korea; \\
\textsuperscript{c} Graduate School of Artificial Intelligence, Pohang University of Science \& Technology, Pohang 37673, Korea;
}}
\cortext[cor1]{Corresponding author. Emails: mailantran@postech.ac.kr (Mai Lan Tran), dongjinlee@postech.ac.kr (Dongjin Lee), jung153@postech.ac.kr (Jae-Hun Jung)}

\begin{abstract}
Common AI music composition algorithms based on artificial neural networks are to train a machine by feeding a large number of music pieces and create artificial neural networks that can produce music similar to the input music data. This approach is a blackbox optimization, that is, the underlying composition algorithm is, in general, not known to users. 

In this paper, we present a way of machine composition that trains a machine the composition principle embedded in the given music data instead of directly feeding music pieces. We propose this approach by using the concept of {\color{black}{Overlap}} matrix proposed in \cite{TPJ}. In \cite{TPJ}, a type of Korean music, so-called the {\it Dodeuri} music such as Suyeonjangjigok  has been analyzed using topological data analysis (TDA), particularly using persistent homology. As the raw music data is not suitable for TDA analysis, the music data is first reconstructed as a graph. The node of the graph is defined as a two-dimensional vector composed of the pitch and duration of each music note. The edge between two nodes is created when those nodes appear consecutively in the music flow. Distance is defined based on the frequency of such appearances. Through TDA on the constructed graph, a unique set of cycles is found for the given music. In \cite{TPJ}, the new concept of the {\it {\color{black}{Overlap}} matrix} has been proposed, which visualizes how those  cycles are interconnected over the music flow, in a matrix form. 

In this paper, we explain how we use the {\color{black}{Overlap}} matrix for machine composition. The {\color{black}{Overlap}} matrix makes it possible to compose a new music piece algorithmically and also  provide a seed music towards the desired artificial neural network. In this paper, we use the {\it Dodeuri} music and explain detailed steps. 

\end{abstract}


\end{frontmatter}

\section{Introduction}
\begin{figure}[h]
    \begin{center}
    \includegraphics[width=0.9\textwidth]{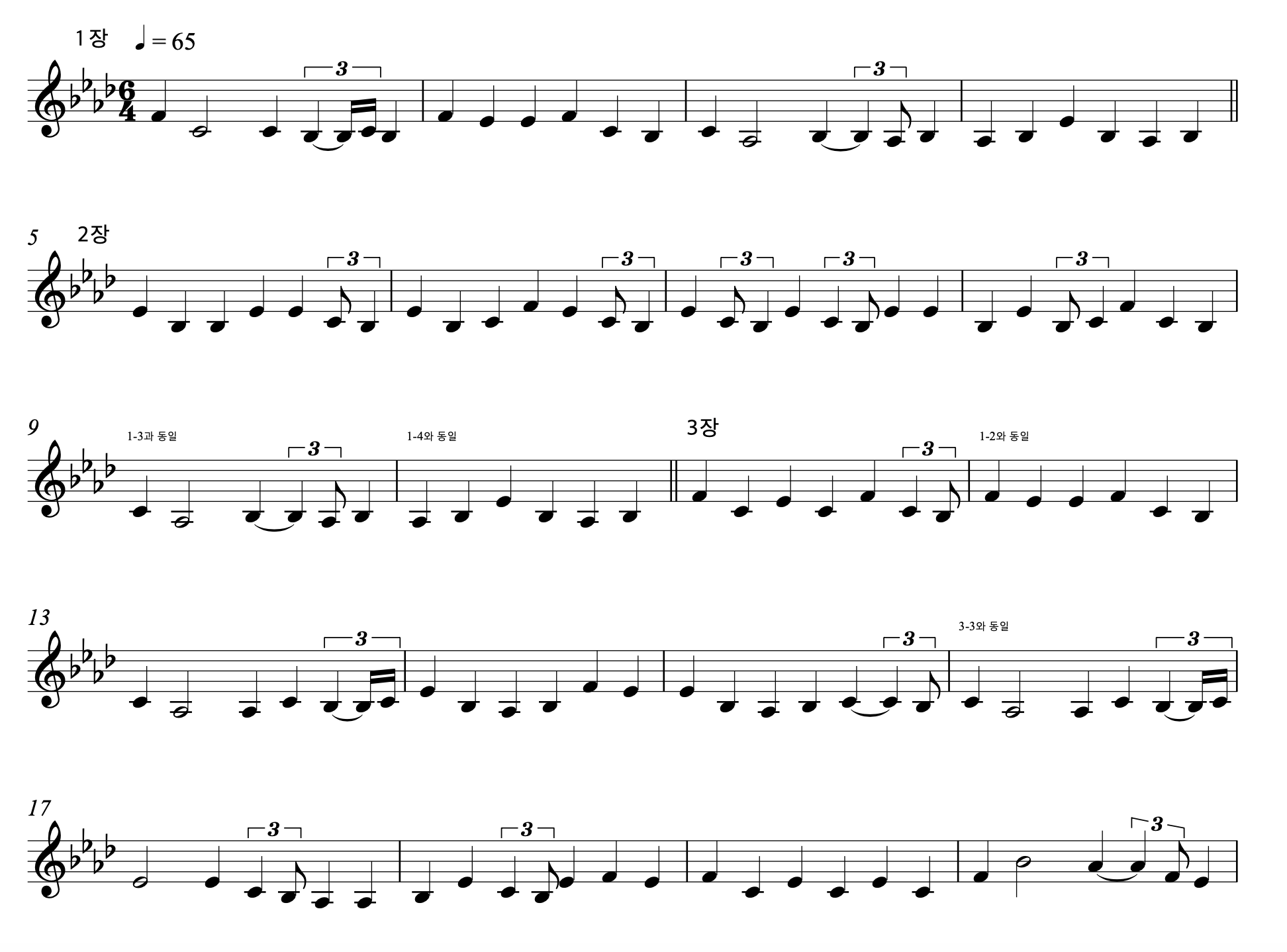}
    \caption{Suyeonjangjigok without ornaments. Credit: Myong-Ok Kim}
    \label{fig:syj_original}
    \end{center}
\end{figure}

Topological data analysis (TDA) has been used in music analysis recently based on persistent homology \cite{Bigo,Bergomi,Bergomithesis}. Persistent homology is an efficient concept for music analysis as it captures the cyclic structures of data \cite{ZC2005,Cohen,EHarer}.  
In \cite{TPJ}, TDA was used to analyze Korean {\it Jung-Ak} music for the first time. {\it Jung-Ak} music\footnote{The literal meaning of {\it Jung Ak} is the {\it right} music.} is a type of music that was played at Royal palaces or among noble communities in old Korea.  {\it Dodeuri} music is one of the most popular Jung-Ak music pieces. As its name indicates, the main characteristics of {\it Dodeuri} (repeat-and-return) music are in its frequent repetition and variation patterns. To analyze such patterns, TDA, particularly persistent homology has been utilized in \cite{TPJ}. Since the raw form of the music is not suitable, the given music is represented as a network \cite{BW,Liu,RCG,LTS10}, for which proper definitions of nodes and edges are provided in \cite{TPJ}. A node is defined as a two-dimensional vector whose first component is the pitch and the second the duration of the music note. If two nodes are placed side-by-side in the music, those two nodes are directly connected and the edge between those two nodes is defined as the connection. The weight of the edge is defined as the frequency of the side-by-side appearance of those two nodes. In such a way, the weight is non-negative integers. In order to apply TDA, the notion of the distance between two nodes is defined as the reciprocal of the edge weight of those two nodes if they are  connected directly. If two nodes are only connected through more than two edges, the distance between those two nodes is defined by the sum of the reciprocals of the weights of edges involved between those two nodes. For the uniqueness, such edges between two nodes are picked in the path that has the smallest number of edges among all possible paths. Once the distance is defined, persistent homology is calculated and the corresponding one-dimensional barcodes are obtained. The one-dimensional barcode contains the one-dimensional hole information \cite{Carlsson}, that is, one-dimensional cycles in the given graph. In \cite{TPJ}, a unique set of total $8$ cycles was found for Suyeonjangjigok in Haegeum instrument -- Suyeonjangjigok (or Suyeonjang in short) is one of the most popular Dodeuri music. 
Figure \ref{fig:syj_original} shows first few lines of Suyeonjangjigok directly translated from Jeongganbo, the old Korean music notation. The version in the figure is a simple version of Suyeonjangjigok without ornaments. Readers can also listen Suyeonjangjigok played with Haegeum instrument in the following YouTube link {\lstinline{https://www.youtube.com/watch?v=_DKo8FjL7Mg&t=461s}} from 0:24 to 5:24. In this paper, we mainly use Suyeonjangjigok as an exemplary for the development of the proposed method. 

In \cite{TPJ}, a new concept of {\it {\color{black}{Overlap}} matrix} of $s$-scale was introduced. The {\color{black}{Overlap}} matrix of $s$-scale, which will be explained in detail in Section \ref{cycles}, is a visualization in a matrix form that shows how those found cycles are interconnected over the music at $s$-scale. It was shown in \cite{TPJ} that the proposed {\color{black}{Overlap}} matrix is useful to understand how the given music is composed and to classify music. In fact, the {\color{black}{Overlap}} matrix explains surprisingly and quantitively well why the Dodeuri music is  different from the Taryong music, a music known as a non Dodeuri music. In this way, the {\color{black}{Overlap}} matrix can be interpreted as the composition algorithm or composition principle of the considered music. The current paper is based on our assumption that the {\color{black}{Overlap}} matrix reveals the composition algorithm of the music considered. Upon such assumption, we propose a way of machine composition using the {\color{black}{Overlap}} matrix. 

Machine composition with Artificial Intelligence (AI) techniques based on artificial neural networks is well known even to non-experts these days \cite{Lopez}. There are various AI composition software packages available as well. AI music composition algorithms based on deep neural networks are to train a machine by feeding music pieces and create artificial neural networks that can produce music similar to the input music data. These approaches are considered as a blackbox optimization. That is, how machine composes with the constructed network is not known to users and the underlying composition algorithm of the generated music pieces is, in general, not explainable. 

In this paper, we present a way of machine composition that trains a machine the composition principle embedded in the given music data. Our proposed method is based on the {\color{black}{Overlap}} matrix explained in the above. As explained in details in Section \ref{cycles}, the {\color{black}{Overlap}} matrix is a kind of visualization method that shows how the key cycles of the music found via TDA are distributed and interconnected one another over the music flow. The main idea of the current paper is to train a machine the composition principle represented by the {\color{black}{Overlap}} matrix in the expectation that the music is algorithmically composed mimicking the input {\color{black}{Overlap}} matrix. One can simply create or design the {\color{black}{Overlap}} matrix and the machine generates the music with such {\color{black}{Overlap}} matrix as a seed music. In this paper, we explain how music can be generated with the {\color{black}{Overlap}} matrix. First, we will explain that the {\color{black}{Overlap}} matrix can be used directly to compose a music algorithmically. Second we will explain how we train a machine the {\color{black}{Overlap}} matrix for building artificial neural networks and generating music. 

The paper is composed of the following sections. In Section \ref{cycles}, we will explain the key elements of the current paper. We first explain the music network and TDA over the constructed network. Then we give detailed mathematical properties of the Overlap matrix.   In Section \ref{nodepool}, we will explain the concept of node pool, which serves as the provider of nodes used for the composition. In Section \ref{algorithm}, we explain how we use the {\color{black}{Overlap}} matrix to generate music algorithmically -- Algorithm A. In Section \ref{ANN}, we explain how we use the Overlap matrix in the context of artificial neural network. First we propose three different methods that can generate the seed music using the {\color{black}{Overlap}} matrix in order to use towards the construction of the artificial neural network. Then, we provide a way to train a machine with the {\color{black}{Overlap}} matrix and construct the corresponding artificial neural networks. In Section \ref{conclusion}, we provide a brief concluding remark and future research questions. 

\section{Cycles and Overlap matrix}
\label{cycles}
In this section we first explain how to construct music network from the given music. The raw form of music is not suitable for TDA. We represent the given music as a graph so that TDA can be applied. As in \cite{TPJ}, we consider Suyeonjangjigok, a Dodeuri type monophonic music written in {\it Jeongganbo}. Jeongganbo is a unique Korean music notation similar to a matrix. Figure \ref{fig:syj_original} shows first few lines of Suyeonjangjigok translated directly from Jeongganbo. For the explanation of reading Jeongganbo, see \cite{TPJ}. Once music network is constructed, we apply TDA and obtain so-called the persistent barcode and the cycles corresponding to the one-dimensional barcode. Based on the barcode and cycles,  we build the {\it {\color{black}{Overlap}} matrix} which will be used together with the cycle information and node frequency distribution for generating new music. We refer readers to \cite{TPJ,Carlsson} for more details of TDA through persistent homology. 

\subsection{Construction of music network}
Consider a monophonic music piece composed of $d$ notes and let $\mathcal{L} = \{n_1, n_2, \ldots, n_d\}$ be the ordered sequence of notes as the music flow. Each note has the information of height (pitch) and length (duration) of the sound to be played, i.e.,
\[
	n_i = (p_i, l_i),
\] 
where $p_i$ is the pitch of note $n_i$ and $l_i$ is its length. Note that by the definition of $n_i$, it is possible that $(p_i,l_i) = (p_j, l_j)$ for $i\neq j$. We then construct the music network $G = (\mathcal{N}, \mathcal{E})$, where $\cal{N}$ is the set of nodes and $\cal{E}$ is the set of edges in $G$. Here, $\cal{N}$ is the set of distinct notes in $\mathcal{L}$, sorted in ascending order in terms of pitch first then length. That is, $\mathcal{N} = \{\nu_1,\nu_2,\ldots,\nu_{q}\}$, where $q\leq d$ is the number of distinct notes in $\mathcal{L}$ and $\nu_j$ has a higher pitch than $\nu_i$ or both have the same pitch but $\nu_j$ has a longer length than $\nu_i$ if $j>i$. We draw an edge between two nodes $\nu_i$ and $\nu_j$, $i\neq j$, if they occur adjacent in time.  
Let $e_{ij} \in \mathcal{E}$ be the edge  whose end points are $\nu_i$ and $\nu_j$. 
The weight or the degree of the edge $e_{ij}$,  $w_{ij}$, between $\nu_i$ and $\nu_j$ is the number of occurrences of those two nodes being adjacent in time. 
For two nodes $\nu_i$ and $\nu_j$ with $i<j$, let $p_{ij}$ be the path with the minimum number of edges between $\nu_i$ and $\nu_j$ found by Dijkstra algorithm. 
The distance between nodes $\nu_i$ and $\nu_j$, $i<j$ is defined to be:
\begin{equation}\label{dij}
	\delta(\nu_i,\nu_j) = \sum \limits_{e_{kl}\in p_{ij}}w_{kl}^{-1}
\end{equation}
where $w_{kl}$ represents the weight of the edge $e_{kl}$ and $p_{ij} = \bigcup e_{kl} $. 
For those music we consider in this paper, since there is no empty Jeonggan where the music is not played, i.e., there is no isolated node, there always exists at least one path between any two distinct nodes $\nu_i$ and $ \nu_j$ even if they do not appear adjacently in the whole music.
Also, it is obvious that $w_{kl}\geq 1$ for any edge $e_{kl}$. Thus, the definition of the distance by \eqref{dij} is well-defined.
Then we form the distance matrix $D = \{\delta_{ij}\}$ as follows:
\[
	\delta_{ij} = \left\{ \begin{array}{lc} 
	                \delta(\nu_i,\nu_j), & i < j \\ 
	                0, & i = j \\
	                \delta_{ji}, & i> j
	                \end{array} \right. 
\]
\subsection{TDA: Barcode and Cycles}
We do not attempt to explain TDA here but refer readers to \cite{Carlsson} if necessary. The graph introduced above is defined with the definition of distance. Consider a point cloud composed of all nodes in $\mathcal{N}$. As all the pair-wise distances between $\nu_i$ and $\nu_j$ are defined, we first build a simplical complex out of the point {\color{black}{cloud}} as a Vietoris-Rips complex to compute persistent homology (see \cite{Carlsson,EHarer}). We note that there are other approaches rather than Vietoris-Rips complex for persistent homology on a graph. Since the main purpose of the current research is to propose a machine composition algorithm, the choice of building algorithm of complexes and filtration method is not critical. Using the distance matrix $D$, we build the corresponding Vietoris-Rips complex and barcode, for which we use the software package Javaplex \cite{AT}. 

\begin{figure}[h]
    \begin{center}
    \includegraphics[width=0.7\textwidth]{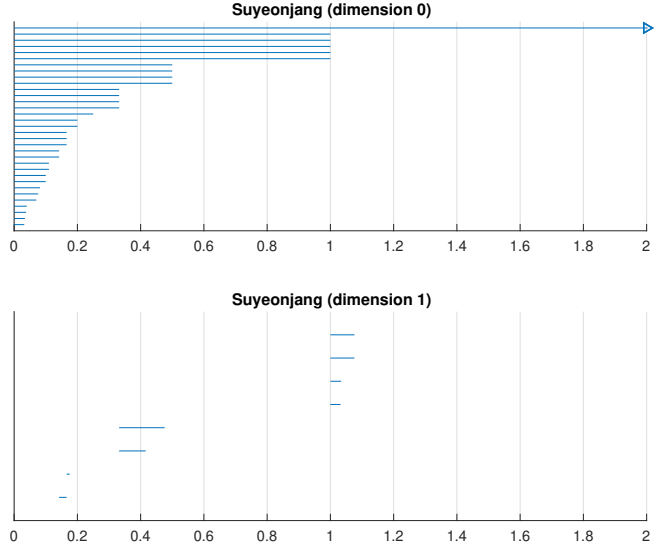}
    \caption{Barcode of Suyeonjang using Vietoris-Rips method. Top: 0-D barcode. Bottom: 1-D barcode. In 1D barcode, we observe 8 non-zero persistence intervals implying 8 cycles in $G$. Credit: \cite{TPJ}}
    \label{fig:syj_barcode}
    \end{center}
\end{figure}

Suyeonjang has total $d = 440$ notes composed of $q = 33$ distinct notes.  \figref{fig:syj_barcode} shows the zero-dimensional (top) and one-dimensional (bottom) barcodes generated by Javaplex applied to $G$ of Suyeonjang. 

In \figref{fig:syj_barcode} the horizontal axis is the filtration value $\tau$. Vertically we have multiple intervals that correspond to generators of the homology groups. In the zeroth dimension we have $33$ generators that correspond to $33$ components when $\tau$ is zero or small, which eventually are connected into a single component when $\tau=1$. The $33$ components are actually those $33$ distinct nodes defined in Suyeonjang. All these components constitute a single component because of the fact that any node in the network connects at least one time with another node, which means that at most when distance $\tau=1$ all nodes in the network are connected. On the other hand, the fact that one component is formed exactly when $\tau = 1$ implies that there exists at least a pair of nodes $\nu_i, \nu_j$ that have distance $\delta (\nu_i, \nu_j)= 1$, i.e., $\nu_i$ and $\nu_j$ are adjacent only once. In the first dimension we see $8$ generators which topologically correspond to eight cycles. It turns out that the interconnection between these cycles is related to the repetition of music melodies known as Dodeuri \cite{TPJ}.

{\color{black}{
For each persistence interval we use the persistence algorithm computing intervals to find a representative cycle. The method \textsf{computeAnnotatedIntervals} in Javaplex is used to find the nodes in the intervals of persistence. In the one dimensional case, the annotated intervals consist of the components in the loops generated in the process of filtration. 
}}

\figref{fig:syj_cycles} shows $8$ Cycles identified by TDA corresponding to $8$ persistence intervals in the one dimensional barcode {\color{black}{of Suyeonjang. We enumerate the cycles by the order of the appearance of their corresponding persistence intervals in the barcode.}} 
That is, the earlier the 1D barcode dies, the lower number is assigned to the corresponding cycle.
For example, the death of Cycle $i$ is earlier than the death of Cycle $j$ if $i < j$. Note that this order is different from the order of their appearance in the actual music {\color{black}{and can be done arbitrarily without affecting the proposed composition algorithms}}.  In the figure, each Cycle is shown with persistence interval, node information including node number, pitch and its length, the latter two of which are encrypted in {\color{black}{the circles filled with different colors and centered by Chinese letters. In fact}}, the Chinese letter in the center of each filled circle corresponds to a specific pitch, and the color of each circle illustrates the node length (see Table \ref{tab:syj_nodeinfo}). The figure also shows edge weight (in normal size in blue), distance between nodes (in small size in blue in brackets) and the average weight (in red in center) which is the simple mean of all edge weights. As shown in the figure, the minimum number of nodes that constitute a cycle is $4$ and the maximum number is $6$. 
The information of corresponding music notes found {\color{black}{in}} all cycles is given in \tabref{tab:syj_nodeinfo}.
{\color{black}{For the purpose of this paper, we will use only the node information of the Cycles $C_i, i=1,\ldots,8$. More precisely, we will use the information of which nodes each Cycle consists of. For example, in the case of Suyeonjang, $C_1=\{ \nu_{18}, \nu_{20}, \nu_{22}, \nu_{27}\}$, $C_2=\{ \nu_{3}, \nu_{6}, \nu_{12}, \nu_{18}\}$ and so on. It should be noted that we do not need detailed information of the names, pitches or lengths of the music notes for building the composition algorithms. Those actual music note information will be used at the finishing stage where we get the generated music for playing.}}
In the next section, we explain in details how the Cycles will be used to construct so-called the Overlap matrix, one of the key ingredients in generating new music.

\include{syj_cycles}


\begin{table}[!h]\centering\resizebox{.7\columnwidth}{!}{%
        \begin{tabular}{c | p{3cm}  p{2cm}  c} 
        \multirow{2}{*}{\makecell{Node symbol\\ in cycles\\(\figref{fig:syj_cycles})}}& 
        \multicolumn{3}{c} {Music note}  \\ 
        \cline{2-4} 
        & Name & Pitch & \makecell{Length \\(Jeonggan)}\\                     \hline\\[-.9em] 
            \begin{tikzpicture}[inner sep=0pt, baseline=(base)] 
            \node[fill=teal!60,draw,shape=circle,inner sep=0pt,minimum size=.6cm] (v) at (0,0.5ex) {\textcolor{black}{\includegraphics[width=0.36cm]{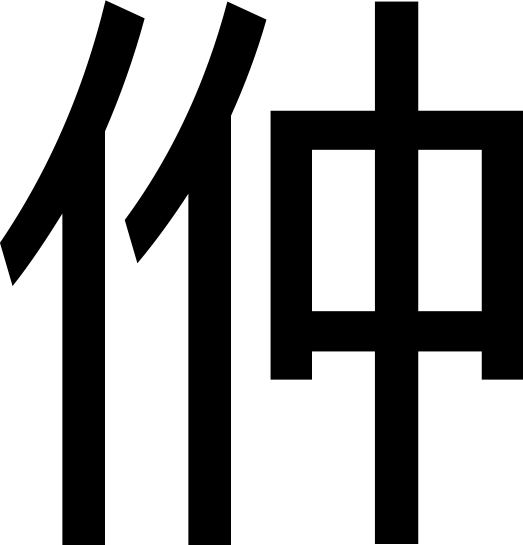}}}; 
            \node (base) at (0,-.5ex) {}; 
            \node at (.7,.05) {$\nu_{0}$}; 
            \node at (1.,0) {${}_{}$}; 
            \end{tikzpicture} 
            & \includegraphics[width=0.36cm]{jung0.png} (Jung) & $G\sharp3$ & 1/3 \\ 
                        \hline\\[-.9em] 
            \begin{tikzpicture}[inner sep=0pt, baseline=(base)] 
            \node[fill=blue!99,draw,shape=circle,inner sep=0pt,minimum size=.6cm] (v) at (0,0.5ex) {\textcolor{white}{\includegraphics[width=0.36cm]{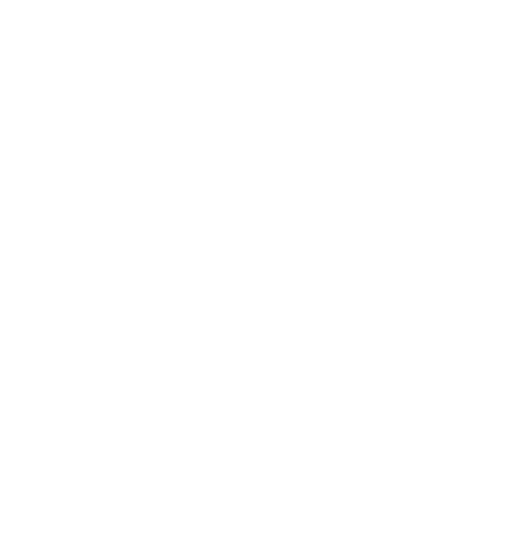}}}; 
            \node (base) at (0,-.5ex) {}; 
            \node at (.7,.05) {$\nu_{1}$}; 
            \node at (1.,0) {${}_{}$}; 
            \end{tikzpicture} 
            & \includegraphics[width=0.36cm]{jung0.png} (Jung) & $G\sharp3$ & 1 \\ 
                        \hline\\[-.9em] 
            \begin{tikzpicture}[inner sep=0pt, baseline=(base)] 
            \node[fill=red!30,draw,shape=circle,inner sep=0pt,minimum size=.6cm] (v) at (0,0.5ex) {\textcolor{black}{\includegraphics[width=0.36cm]{jung0.png}}}; 
            \node (base) at (0,-.5ex) {}; 
            \node at (.7,.05) {$\nu_{2}$}; 
            \node at (1.,0) {${}_{}$}; 
            \end{tikzpicture} 
            & \includegraphics[width=0.36cm]{jung0.png} (Jung) & $G\sharp3$ & 2 \\ 
                        \hline\\[-.9em] 
            \begin{tikzpicture}[inner sep=0pt, baseline=(base)] 
            \node[fill=teal!60,draw,shape=circle,inner sep=0pt,minimum size=.6cm] (v) at (0,0.5ex) {\textcolor{black}{\includegraphics[width=0.36cm]{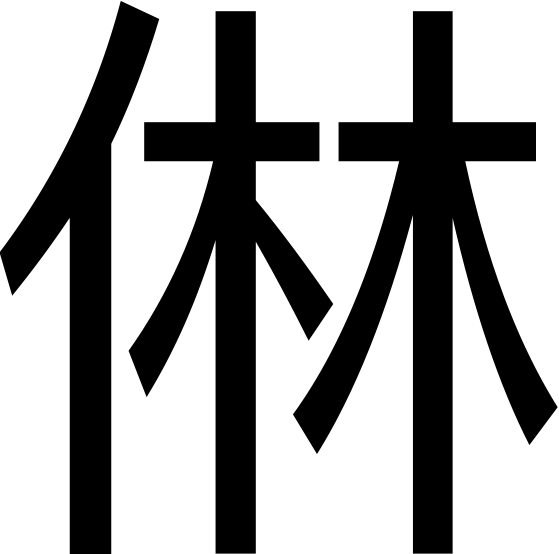}}}; 
            \node (base) at (0,-.5ex) {}; 
            \node at (.7,.05) {$\nu_{3}$}; 
            \node at (1.,0) {${}_{}$}; 
            \end{tikzpicture} 
            & \includegraphics[width=0.36cm]{Im0.png} (Im) & $A\sharp3$ & 1/3 \\ 
                        \hline\\[-.9em] 
            \begin{tikzpicture}[inner sep=0pt, baseline=(base)] 
            \node[fill=blue!99,draw,shape=circle,inner sep=0pt,minimum size=.6cm] (v) at (0,0.5ex) {\textcolor{white}{\includegraphics[width=0.36cm]{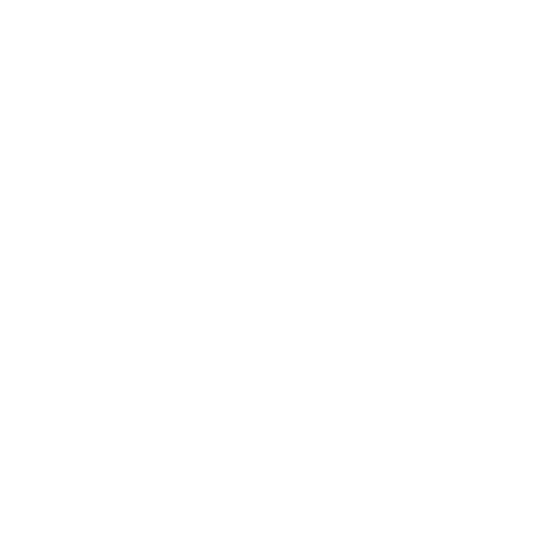}}}; 
            \node (base) at (0,-.5ex) {}; 
            \node at (.7,.05) {$\nu_{6}$}; 
            \node at (1.,0) {${}_{}$}; 
            \end{tikzpicture} 
            & \includegraphics[width=0.36cm]{Im0.png} (Im) & $A\sharp3$ & 1 \\ 
                        \hline\\[-.9em] 
            \begin{tikzpicture}[inner sep=0pt, baseline=(base)] 
            \node[fill=orange!99,draw,shape=circle,inner sep=0pt,minimum size=.6cm] (v) at (0,0.5ex) {\textcolor{white}{\includegraphics[width=0.36cm]{Imw.png}}}; 
            \node (base) at (0,-.5ex) {}; 
            \node at (.7,.05) {$\nu_{7}$}; 
            \node at (1.,0) {${}_{}$}; 
            \end{tikzpicture} 
            & \includegraphics[width=0.36cm]{Im0.png} (Im) & $A\sharp3$ & 5/3 \\ 
                        \hline\\[-.9em] 
            \begin{tikzpicture}[inner sep=0pt, baseline=(base)] 
            \node[fill=blue!99,draw,shape=circle,inner sep=0pt,minimum size=.6cm] (v) at (0,0.5ex) {\textcolor{white}{\includegraphics[width=0.36cm]{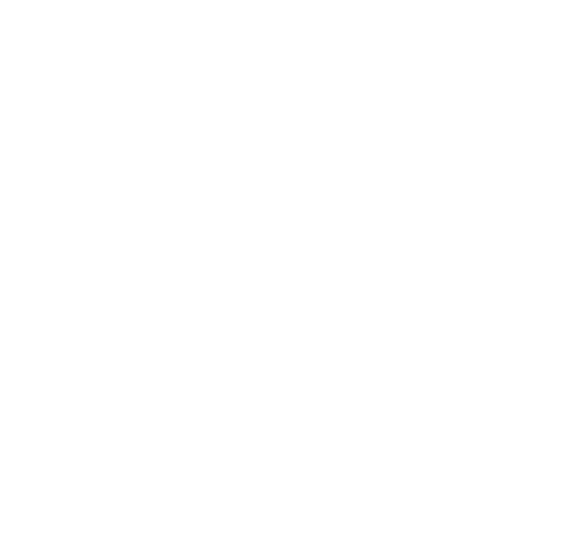}}}; 
            \node (base) at (0,-.5ex) {}; 
            \node at (.7,.05) {$\nu_{11}$}; 
            \node at (1.,0) {${}_{}$}; 
            \end{tikzpicture} 
            & \includegraphics[width=0.36cm]{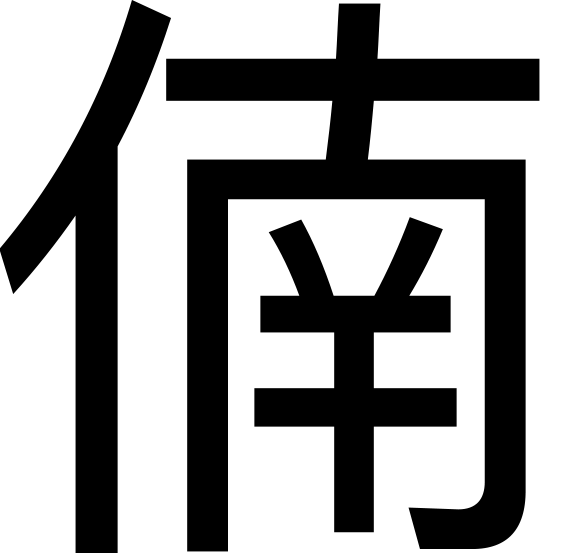} (Nam) & $C4$ & 1 \\ 
                        \hline\\[-.9em] 
            \begin{tikzpicture}[inner sep=0pt, baseline=(base)] 
            \node[fill=orange!99,draw,shape=circle,inner sep=0pt,minimum size=.6cm] (v) at (0,0.5ex) {\textcolor{white}{\includegraphics[width=0.36cm]{namw.png}}}; 
            \node (base) at (0,-.5ex) {}; 
            \node at (.7,.05) {$\nu_{12}$}; 
            \node at (1.,0) {${}_{}$}; 
            \end{tikzpicture} 
            & \includegraphics[width=0.36cm]{nam0.png} (Nam) & $C4$ & 5/3 \\ 
                        \hline\\[-.9em] 
            \begin{tikzpicture}[inner sep=0pt, baseline=(base)] 
            \node[fill=teal!60,draw,shape=circle,inner sep=0pt,minimum size=.6cm] (v) at (0,0.5ex) {\textcolor{black}{黃}}; 
            \node (base) at (0,-.5ex) {}; 
            \node at (.7,.05) {$\nu_{16}$}; 
            \node at (1.,0) {${}_{}$}; 
            \end{tikzpicture} 
            & 黃 (Hwang) & $D\sharp4$ & 1/3 \\ 
                        \hline\\[-.9em] 
            \begin{tikzpicture}[inner sep=0pt, baseline=(base)] 
            \node[fill=blue!99,draw,shape=circle,inner sep=0pt,minimum size=.6cm] (v) at (0,0.5ex) {\textcolor{white}{黃}}; 
            \node (base) at (0,-.5ex) {}; 
            \node at (.7,.05) {$\nu_{18}$}; 
            \node at (1.,0) {${}_{}$}; 
            \end{tikzpicture} 
            & 黃 (Hwang) & $D\sharp4$ & 1 \\ 
                        \hline\\[-.9em] 
            \begin{tikzpicture}[inner sep=0pt, baseline=(base)] 
            \node[fill=teal!60,draw,shape=circle,inner sep=0pt,minimum size=.6cm] (v) at (0,0.5ex) {\textcolor{black}{太}}; 
            \node (base) at (0,-.5ex) {}; 
            \node at (.7,.05) {$\nu_{20}$}; 
            \node at (1.,0) {${}_{}$}; 
            \end{tikzpicture} 
            & 太 (Tae) & $F4$ & 1/3 \\ 
                        \hline\\[-.9em] 
            \begin{tikzpicture}[inner sep=0pt, baseline=(base)] 
            \node[fill=blue!30,draw,shape=circle,inner sep=0pt,minimum size=.6cm] (v) at (0,0.5ex) {\textcolor{black}{太}}; 
            \node (base) at (0,-.5ex) {}; 
            \node at (.7,.05) {$\nu_{21}$}; 
            \node at (1.,0) {${}_{}$}; 
            \end{tikzpicture} 
            & 太 (Tae) & $F4$ & 2/3 \\ 
                        \hline\\[-.9em] 
            \begin{tikzpicture}[inner sep=0pt, baseline=(base)] 
            \node[fill=blue!99,draw,shape=circle,inner sep=0pt,minimum size=.6cm] (v) at (0,0.5ex) {\textcolor{white}{太}}; 
            \node (base) at (0,-.5ex) {}; 
            \node at (.7,.05) {$\nu_{22}$}; 
            \node at (1.,0) {${}_{}$}; 
            \end{tikzpicture} 
            & 太 (Tae) & $F4$ & 1 \\ 
                        \hline\\[-.9em] 
            \begin{tikzpicture}[inner sep=0pt, baseline=(base)] 
            \node[fill=orange!99,draw,shape=circle,inner sep=0pt,minimum size=.6cm] (v) at (0,0.5ex) {\textcolor{white}{太}}; 
            \node (base) at (0,-.5ex) {}; 
            \node at (.7,.05) {$\nu_{23}$}; 
            \node at (1.,0) {${}_{}$}; 
            \end{tikzpicture} 
            & 太 (Tae) & $F4$ & 5/3 \\ 
                        \hline\\[-.9em] 
            \begin{tikzpicture}[inner sep=0pt, baseline=(base)] 
            \node[fill=red!30,draw,shape=circle,inner sep=0pt,minimum size=.6cm] (v) at (0,0.5ex) {\textcolor{black}{太}}; 
            \node (base) at (0,-.5ex) {}; 
            \node at (.7,.05) {$\nu_{24}$}; 
            \node at (1.,0) {${}_{}$}; 
            \end{tikzpicture} 
            & 太 (Tae) & $F4$ & 2 \\ 
                        \hline\\[-.9em] 
            \begin{tikzpicture}[inner sep=0pt, baseline=(base)] 
            \node[fill=teal!60,draw,shape=circle,inner sep=0pt,minimum size=.6cm] (v) at (0,0.5ex) {\textcolor{black}{仲}}; 
            \node (base) at (0,-.5ex) {}; 
            \node at (.7,.05) {$\nu_{25}$}; 
            \node at (1.,0) {${}_{}$}; 
            \end{tikzpicture} 
            & 仲 (Jung) & $G\sharp4$ & 1/3 \\ 
                        \hline\\[-.9em] 
            \begin{tikzpicture}[inner sep=0pt, baseline=(base)] 
            \node[fill=blue!30,draw,shape=circle,inner sep=0pt,minimum size=.6cm] (v) at (0,0.5ex) {\textcolor{black}{仲}}; 
            \node (base) at (0,-.5ex) {}; 
            \node at (.7,.05) {$\nu_{26}$}; 
            \node at (1.,0) {${}_{}$}; 
            \end{tikzpicture} 
            & 仲 (Jung) & $G\sharp4$ & 2/3 \\ 
                        \hline\\[-.9em] 
            \begin{tikzpicture}[inner sep=0pt, baseline=(base)] 
            \node[fill=blue!99,draw,shape=circle,inner sep=0pt,minimum size=.6cm] (v) at (0,0.5ex) {\textcolor{white}{仲}}; 
            \node (base) at (0,-.5ex) {}; 
            \node at (.7,.05) {$\nu_{27}$}; 
            \node at (1.,0) {${}_{}$}; 
            \end{tikzpicture} 
            & 仲 (Jung) & $G\sharp4$ & 1 \\ 
                        \hline\\[-.9em] 
            \begin{tikzpicture}[inner sep=0pt, baseline=(base)] 
            \node[fill=red!30,draw,shape=circle,inner sep=0pt,minimum size=.6cm] (v) at (0,0.5ex) {\textcolor{black}{仲}}; 
            \node (base) at (0,-.5ex) {}; 
            \node at (.7,.05) {$\nu_{29}$}; 
            \node at (1.,0) {${}_{}$}; 
            \end{tikzpicture} 
            & 仲 (Jung) & $G\sharp4$ & 2 \\ 
                        \hline\\[-.9em] 
            \begin{tikzpicture}[inner sep=0pt, baseline=(base)] 
            \node[fill=blue!30,draw,shape=circle,inner sep=0pt,minimum size=.6cm] (v) at (0,0.5ex) {\textcolor{black}{林}}; 
            \node (base) at (0,-.5ex) {}; 
            \node at (.7,.05) {$\nu_{30}$}; 
            \node at (1.,0) {${}_{}$}; 
            \end{tikzpicture} 
            & 林 (Im) & $A\sharp4$ & 2/3 \\ 
            \end{tabular}}\caption{Information of all nodes that appear in cycles identified by TDA in  Suyeonjang. All nodes are listed in ascending order in terms of pitch.  There are all $20$ nodes that appear in any of $8$ Cycles. The table shows the music note, pitch and length (Table 2 in \cite {TPJ}).  }
            \label{tab:syj_nodeinfo}
\end{table}


\subsection{Overlap matrices} \label{sec:om}
In \cite{TPJ} the {\color{black}{Overlap}} matrix was introduced. In this section, we provide a formal definition of the {\color{black}{Overlap}} matrix and its mathematical properties. 

From now on, let $s$ be a positive integer.
\definition{A \textit{binary matrix} is a matrix whose entries are either $0$ or $1$.
}
\definition{ A binary matrix $M_{k\times d}^s=\{m_{ij}^s\}$ is said to \textit{belong to $s$-scale} if for all $i = 1,\ldots,k$ we have that $m_{ij}^s = 1$ if and only if there exist nonnegative integers $t,l$ satisfying $t+l \geq s-1$ such that 
\begin{equation*}
	m_{i,j-l}^s = m_{i,j-l-1}^s = \ldots = m_{ij}^s = \ldots = m_{i,j+t-1}^s = m_{i,j+t}^s = 1.
\end{equation*}	
}
Notice that  there are $t+l+1$ entries from $m_{i,j-l}^s$ to $m_{i,j+t}^s$:
\begin{center}
\begin{tikzpicture}[scale=1.2]
    \draw (0,0) -- (9,0);
    \node[circle,fill=black,inner sep=0pt,minimum size=5pt,label=below:{$m_{i,j-l}^s$}] at (1,0) {};
    \node[circle,fill=black,inner sep=0pt,minimum size=5pt,label=above:{$m_{i,j-l-1}^s$}] at (2,0) {};
    \node[circle,fill=black,inner sep=0pt,minimum size=5pt,label=below:{$\ldots$}] at (3,0) {};
    \node[circle,fill=black,inner sep=0pt,minimum size=5pt,label=below:{$m_{i,j}^s$}] at (4.5,0) {};
    \node[circle,fill=black,inner sep=0pt,minimum size=5pt,label=below:{$\ldots$}] at (5.5,0) {};
    \node[circle,fill=black,inner sep=0pt,minimum size=5pt,label=above:{$m_{i,j+t-1}^s$}] at (7,0) {};
    \node[circle,fill=black,inner sep=0pt,minimum size=5pt,label=below:{$m_{i,j+t}^s$}] at (8,0) {};
    \node at (4.5,-1){$\underbrace{\phantom{n + 1\text{ length of underbrace longer longer longer}}}_{t+l+1 \text{ entries }}$};
\end{tikzpicture}
 \end{center}
Thus, a binary matrix  $M_{k\times d}^s=\{m_{ij}^s\}$ belongs to $s$-scale if and only if on each row of $M_{k\times d}^s$ any entry equal to $1$ should be staying in a consecutive sequence of length at least $s$ columns that equal to $1$.

Unless other mentioned, let $\mathcal{O}$ be a music piece composed of $d$ notes that flows in the following order $\mathcal{L} = \{ n_1, \ldots, n_d\}$ and assume that the barcode for it in the first dimension consists of $k$ generators which topologically correspond to $k$ Cycles, $C_1, \ldots, C_k$. We define the binary and integer Overlap matrices {\color{black}{of}} $s$-scale for $\mathcal{O}$ as follows.

\definition{Matrix $M_{k\times d}^s=\{m_{ij}^s\}$ is called the \textit{binary Overlap matrix} {\color{black}{of}} $s$-scale for $\mathcal{O}$ if it satisfies the following conditions 
\[
m_{ij}^s = \begin{cases}
	1, & \text{ if } \exists~ t,l \geq 0 \text{ satisfying } t + l \geq s-1 \text{ such that }\\ 	    
	& \hspace{0.5cm}   n_{j-l}, n_{j-l-1}, \ldots, n_j, \ldots, n_{j+t-1}, n_{j+t} \in C_i,\\
	0, & \text{ otherwise,}
	\end{cases}
\]
for all $i = 1,\ldots, k$; $j = 1,\ldots,d$.
}\label{def:bom}

\definition{
Matrix $M_{k\times d}^s=\{m_{ij}^s\}$ is called the \textit{integer Overlap matrix} {\color{black}{of}} $s$-scale for $\mathcal{O}$ if it satisfies the following conditions 
\[
m_{ij}^s = \begin{cases}
	n_j, & \text{ if } \exists~ t,l \geq 0 \text{ satisfying } t + l \geq s-1 \text{ such that }\\ 	    
	& \hspace{0.5cm}   n_{j-l}, n_{j-l-1}, \ldots, n_j, \ldots, n_{j+t-1}, n_{j+t} \in C_i,\\
	0, & \text{ otherwise,}
	\end{cases}
\]
for all $i = 1,\ldots, k$; $j = 1,\ldots,d$.
}\label{def:iom}

%
%
\remark{
Given the integer Overlap matrix {\color{black}{of}} $s$-scale for a music $\mathcal{O}$, its corresponding binary Overlap matrix is uniquely determined and easily obtained by replacing nonzero entries in the integer Overlap matrix with $1$. The converse is not true.
} \label{rmk:iom_to_bom}

\begin{proposition} \label{prop}
If $M_{k\times d}^s=\{m_{ij}^s\}$ is the binary Overlap matrix {\color{black}{of}} $s$-scale for $\mathcal{O}$, then the followings hold
\begin{enumerate}[(i)]
	\item $M_{k\times d}^s$ is a binary matrix belonging to $s$-scale.
	\item $m_{ij}^s = 1$ implies that $n_j \in C_i$.
	\item If $n_j \notin C_i$ then $m_{ij}^s = 0$.
\end{enumerate}
\end{proposition}
\begin{remark}
The converse of $(ii)$ and $(iii)$ is not necessarily true.
\end{remark}
\begin{proof}[Proof of Proposition \ref{prop}]
Let  $M_{k\times d}^s=\{m_{ij}^s\}$ be the binary Overlap matrix {\color{black}{of}} $s$-scale for $\mathcal{O}$. It is easy to see that $(ii)$ and $(iii)$ are straightforward from the definition \ref{def:bom}. To prove $(i)$, since $m_{ij}^s$ is either $0$ or $1$ for all $i,j$, hence $M_{k\times d}^s$ is a binary matrix, it remains to show that on each row of $M_{k\times d}^s$ any entry equal to $1$ stays in a consecutive sequence of length at least $s$ columns that equal to 1.

Let $m_{ij}^s=1$. By definition  \ref{def:bom}, there exist $ t,l \geq 0 $ satisfying $t + l \geq s-1$ such that $n_{j-l}$, $n_{j-l-1}, \ldots, n_j, \ldots, n_{j+t-1}$, $n_{j+t} \in C_i$. In other words, there exists a consecutive sequence of at least $s$ notes including $n_j$ in $\mathcal{O}$ that belong to $C_i$. Now, in turn, $m_{i,j-l}^s = 1$ since there exist $u=0$, $r = t+l \geq0$ satisfying $u+r = t+l \geq s-1$ such that $n_{j-l-u} (= n_{j-l}), n_{j-l-u-1} (=n_{j-l-1}) \ldots, n_{j-l+r-1} ( = n_{j+t-1}), n_{j-l+r} (= n_{j+t}) \in C_i$. Analogously, we can show that $m_{i,j-l-1}^s = \ldots = m_{i,j+t}^s = 1$. Thus, $M_{k\times d}^s=\{m_{ij}^s\}$ belongs to $s$-scale.
\end{proof}
\begin{remark} Given a music piece $\mathcal{O}$, the integer Overlap matrix (and thus the binary Overlap matrix as well, by Remark \ref{rmk:iom_to_bom}) {\color{black}{of  $s$-scale $M_{k\times d}^s$ for $\mathcal{O}$}} is uniquely determined. 
\end{remark}
In Algorithm \ref{algorithm1} we give an algorithm to compute the integer Overlap matrix {\color{black}{of}} $s$-scale $M_{k\times d}^s$ for a given music $\mathcal{O}$. 

\algdef{SE}[SUBALG]{Indent}{EndIndent}{}{\algorithmicend\ }%
\algtext*{Indent}
\algtext*{EndIndent}
\begin{algorithm}
\caption{Algorithm to compute integer Overlap matrix {\color{black}{of}} $s$-scale $M_{k\times d}^s$}\label{algorithm1}
\begin{algorithmic}
\State Given $\mathcal{O}$ and Cycle information $C_1, \ldots, C_k$.
\State Set $M_{k\times d}^s = \mathbb{O}_{k\times d}$ (zero matrix).
\State Let $j = 1$. 
\State For each row $i = 1,\ldots, k$, repeat the followings until $j=d$. 
\State {\bf Step 1: Find}  
\begin{equation}
q = \argmin_\beta \{ j\leq \beta \leq d: n_\beta \in C_i\}.
\label{inCi}
\end{equation}
\Indent\Indent
		 \If {\eqref{inCi} has no solution} 
			\State break
		\Else
			\State Go to Step 2.
		\EndIf
\EndIndent
\EndIndent
\State {\bf Step 2: Find} 
\begin{equation}
\label{not_inCi}
r = \argmin_\gamma \{ q < \gamma \leq d: n_\gamma \notin C_i\}.
\end{equation}
\Indent\Indent
	\If { \eqref{not_inCi} has no solution} 
		\If { $d-q \geq s-1$}
 			$$m_{i,j}^s = n_j, \quad j = q,\ldots,d.$$
		\EndIf
         \State {\bf break}         
         \ElsIf {\eqref{not_inCi} has a solution and $r-q\geq s$} 
   		$$m_{i,j}^s = n_j, \quad j = q, \ldots r-1.$$
	\EndIf
\EndIndent
\EndIndent
\State Set $j = r$ and come back to Step 1.
\end{algorithmic}
\end{algorithm}

\begin{figure}[!h]
	\begin{center}
            	\includegraphics[scale=0.6]{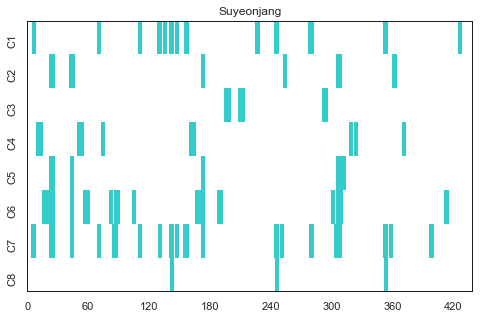}
		\caption{The binary Overlap matrix {\color{black}{of $4$-scale for Suyeonjang}}. The zero entries are left blank and the entries that equal to $1$ are colored. {\color{black}{Credit: \cite{TPJ}}} }
		\label{fig:syj_overlap4}
	\end{center}
\end{figure}

In the case of Suyeonjang which is composed of $d=440$ notes and has in total eight Cycles, the binary Overlap matrix {\color{black}{of}} $4$-scale  $M_{8\times440}^4$ is displayed in \figref{fig:syj_overlap4}. In \figref{fig:syj_overlap4} the horizontal axis represents the time sequence the music flows and the vertical axis represents the cycle number, from $C_1$ to $C_8$.  The zero entries are left blank and the entries that equal to $1$ are colored. Notice that {\color{black}{at}} $s$-scale, each colored block is of length at least $s$ entries.

According to the definition of the binary Overlap matrix $M_{k\times d}^s$, the entries $m_{ij}^s$ can be either $0$ or $1$ depending on whether there exists a consecutive sequence of at least $s$ notes containing the note $n_j$ of the music that belongs to the Cycle $C_i$ or not. 
A zero entry $m_{ij}^s=0$ does not necessarily mean that the note $n_j$ does not belong to the Cycle $C_i$. It can be the case that the note $n_j$ belongs to the Cycle $C_i$ but the consecutive sequence of notes containing note $n_j$ belonging to Cycle $C_i$ is not long enough on the scale being considered. 
On the other hand, if $m_{ij}^s=1$ we can say for sure that the note $n_j$ of the music belongs to the Cycle $C_i$. 
Thus, 
the $j$-th column of the matrix $M_{k\times n}^s$ provide the information of how many cycles, as well as which ones, are overlapping ``{\color{black}{at}} $s$-scale'' at this point. 
This is indeed the motivation why we call it the Overlap matrix.
For example, let us take a close look at the binary Overlap matrix $M_{8\times 440}^4$ {\color{black}{of}} $4$-scale for Suyeonjang. The first column of $M_{8\times 440}^4$ which is
\begin{equation*}
	\begin{bmatrix}
		0 & 0 & 0 & 0 & 0 & 0 & 0 & 0
	\end{bmatrix}^T
\end{equation*}
can mean that the first node $n_1$ does not belong to any Cycle, or in fact it does belong to some Cycle but at least one of the notes $n_2, n_3, n_4$ does not belong to that Cycle. On the other hands, the $25$-th column of $M_{8\times 440}^4$ which is
\begin{equation*}
	\begin{bmatrix}
		0 & 1 & 0 & 0 & 1 & 1 & 1 & 0
	\end{bmatrix}^T
\end{equation*}
implies that the $25$-th note at least belongs to 4 Cycles, that are $C_2$, $C_5$, $C_6$, $C_7$. Similar to the first note $n_1$, it is inconclusive whether or not the $25$-th note belongs to $C_1, C_3, C_4, C_8$.

\vspace{0.3cm}
\begin{minipage}{.5\textwidth}
\begin{center}
\text{first column} \\ \vspace{0.5cm}
\scalebox{0.6}{
\begin{tikzpicture}[rotate=-90,every node/.style={minimum size=1cm-\pgflinewidth, outer sep=0pt}] 
    \draw[step=1cm,color=black] (0,0) grid (8,1);
\end{tikzpicture}
}
\end{center}
\end{minipage}%
\begin{minipage}{.3\textwidth}
\begin{center}
\text{$25$-th column} \\ \vspace{0.5cm}
\scalebox{0.6}{
\begin{tikzpicture}[rotate=-90,every node/.style={minimum size=1cm-\pgflinewidth, outer sep=0pt}] 
    \draw[step=1cm,color=black] (0,0) grid (8,1);
    \node[fill={rgb:cyan,0.7; green,0.2; blue,0.2}] at (1.5,0.5) {$C_2$};
    \node[fill={rgb:cyan,0.7; green,0.2; blue,0.2}] at (4.5,0.5) {$C_5$};
    \node[fill={rgb:cyan,0.7; green,0.2; blue,0.2}] at (5.5,0.5) {$C_6$};
    \node[fill={rgb:cyan,0.7; green,0.2; blue,0.2}] at (6.5,0.5) {$C_7$};
\end{tikzpicture}
}
\end{center}
\end{minipage}
\vspace{0.5cm}
\definition {A Cycle $C_i$ is said to \textit{survive} at note $n_j$ {\color{black}{at}} $s$-scale if $m_{ij}^s = 1$.
}

Denote by $S_j$ the set of Cycles which survive at note $n_j$ {\color{black}{at}} $s$-scale. It is obvious that
\begin{equation} \label{eq:S}
	S_j = \begin{cases}
		\emptyset, & \text { if } m_{ij}^s = 0, \forall i = 1,\ldots,k, \\ 
		\{ C_t | t \in I_j \}, & \text { if } m_{ij}^s = \chi_{I_j} (i),
	\end{cases}
\end{equation}
for $j = 1, \ldots, d$. Here, $I_j$ is an index set, $I_j \subset \{ 1, 2, \ldots, k\}$ and $\chi_{I_j}$ is the indicator function $\chi_{I_j}: \{ 1, 2, \ldots, k\} \to \{0,1\}$ such that
\begin{equation*}
	\chi_{I_j}(x) = \begin{cases}
	1, & \text{ if } x \in {I_j},\\
	0, & \text{ if } x \notin {I_j}.
	\end{cases}	
\end{equation*}
Indeed, $I_j$ is the set of all the indices $i$ where $m_{ij}^s=1$ for a given $j$. In the case of Suyeonjang we have, for example, that $S_1 = \emptyset$ and $S_{25} = \{C_2, C_5, C_6, C_7\}$.

\section{Node pool}
\label{nodepool}
\begin{figure}[!h]
	\begin{center}
		\includegraphics[scale=0.8]{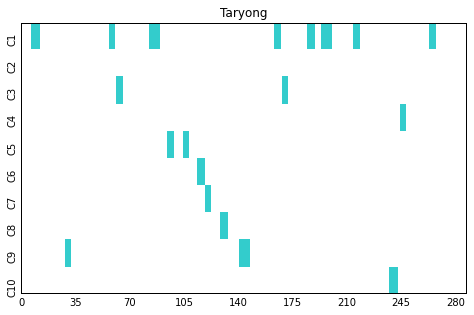}
		\caption{The binary Overlap matrix {\color{black}{of $4$-scale for Taryong}}. The zero entries are left blank and the entries that equal to $1$ are colored. Credit: \cite{TPJ}  } 
		\label{fig:tr_overlap4}
	\end{center}
\end{figure}

First consider the algorithmic composition by following the pattern of the binary Overlap matrix {\color{black}{of}} given $s$-scale of the considered music. Notice that {\color{black}{at}} $s$-scale the Cycles obtained from the considered music by TDA tools only overlap at certain notes (see the binary Overlap matrix of Suyeonjang in \figref{fig:syj_overlap4}) or sometimes they do not overlap at all (see the binary Overlap matrix of {\color{black}{a music called Taryong \cite{TPJ}}} in \figref{fig:tr_overlap4}). Also, there are many notes along the music flow where there is no Cycle surviving {\color{black}{at}} $s$-scale. At those notes, more freedom of node choice can be given. We will build up a so-called Node pool (denoted by $\mathcal{P}$), from which we choose node for those places where there is no Cycle surviving {\color{black}{at}} the considered $s$-scale.  The Node pool is a collection of nodes that satisfies the node frequency distribution.  
\begin{table}[!h]
  \begin{center}
    \caption{Frequency versus node for Suyeonjang, Songkuyeo and Taryong music. Credit: \cite{TPJ}.}
    \label{tab:freq}
    {\footnotesize{
    \begin{tabular}{| c | |c c|| c c|| c c|} 
          \hline
    {Rank}&   {Suyeonjang} &  & {Songkuyeo} &  & {Taryong} &  \\
      \hline
1 & $n_{18}$ & 76 & $n_{20}$ & 65 & $n_{16}$ & 38 \\ 
2 & $n_{6}$ & 57 & $n_{31}$ & 53 & $n_{11}$ & 28 \\ 
3 & $n_{11}$ & 44 & $n_{13}$ & 45 & $n_{13}$ & 23 \\ 
4 & $n_{22}$ & 44 & $n_{26}$ & 44 & $n_{26}$ & 18 \\ 
5 & $n_{1}$ & 30 & $n_{8}$ & 27 & $n_{29}$ & 17 \\ 
6 & $n_{20}$ & 26 & $n_{18}$ & 23 & $n_{31}$ & 15 \\ 
7 & $n_{27}$ & 22 & $n_{4}$ & 18 & $n_{28}$ & 15 \\ 
8 & $n_{3}$ & 16 & $n_{33}$ & 18 & $n_{3}$ & 14 \\ 
9 & $n_{28}$ & 14 & $n_{6}$ & 11 & $n_{18}$ & 13 \\ 
10 & $n_{12}$ & 10 & $n_{16}$ & 11 & $n_{15}$ & 11 \\ 
11 & $n_{16}$ & 9 & $n_{25}$ & 11 & $n_{12}$ & 10 \\ 
12 & $n_{26}$ & 9 & $n_{19}$ & 10 & $n_{22}$ & 10 \\ 
13 & $n_{31}$ & 9 & $n_{24}$ & 10 & $n_{6}$ & 9 \\ 
14 & $n_{2}$ & 7 & $n_{27}$ & 10 & $n_{32}$ & 8 \\ 
15 & $n_{4}$ & 7 & $n_{28}$ & 9 & $n_{17}$ & 7 \\ 
16 & $n_{23}$ & 7 & $n_{32}$ & 8 & $n_{20}$ & 7 \\ 
17 & $n_{9}$ & 6 & $n_{2}$ & 6 & $n_{4}$ & 5 \\ 
18 & $n_{10}$ & 6 & $n_{15}$ & 5 & $n_{9}$ & 4 \\ 
19 & $n_{5}$ & 5 & $n_{7}$ & 4 & $n_{0}$ & 3 \\ 
20 & $n_{8}$ & 5 & $n_{11}$ & 3 & $n_{14}$ & 3 \\ 
21 & $n_{13}$ & 5 & $n_{12}$ & 3 & $n_{2}$ & 2 \\ 
22 & $n_{0}$ & 4 & $n_{14}$ & 3 & $n_{5}$ & 2 \\ 
23 & $n_{7}$ & 3 & $n_{17}$ & 3 & $n_{7}$ & 2 \\ 
24 & $n_{17}$ & 3 & $n_{21}$ & 3 & $n_{8}$ & 2 \\ 
25 & $n_{19}$ & 3 & $n_{23}$ & 3 & $n_{19}$ & 2 \\ 
26 & $n_{21}$ & 2 & $n_{35}$ & 3 & $n_{21}$ & 2 \\ 
27 & $n_{25}$ & 2 & $n_{0}$ & 2 & $n_{27}$ & 2 \\ 
28 & $n_{29}$ & 2 & $n_{3}$ & 2 & $n_{33}$ & 2 \\ 
29 & $n_{30}$ & 2 & $n_{9}$ & 2 & $n_{34}$ & 2 \\ 
30 & $n_{32}$ & 2 & $n_{10}$ & 2 & $n_{35}$ & 2 \\ 
31 & $n_{14}$ & 1 & $n_{36}$ & 2 & $n_{1}$ & 1 \\ 
32 & $n_{15}$ & 1 & $n_{34}$ & 2 & $n_{10}$ & 1 \\ 
33 & $n_{24}$ & 1 & $n_{1}$ & 1 & $n_{23}$ & 1 \\ 
34 & & & $n_{5}$ & 1 & $n_{24}$ & 1 \\ 
35 & & & $n_{22}$ & 1 & $n_{25}$ & 1 \\
36 & & & $n_{29}$ & 1 & $n_{30}$ & 1 \\
37 & & & $n_{30}$ & 1 & $n_{38}$ & 1 \\
38 & & &  & & $n_{36}$ & 1 \\ 
39 & & &  & & $n_{37}$ & 1 \\
40 & & &  & & $n_{39}$ & 1 \\ 
 \hline
       \end{tabular}
       }}
  \end{center}
\end{table}
Let us take Suyeonjang as an example again. According to our node definition, Suyeonjang is of length $440$ notes that consists of $33$ distinct nodes. The node frequency distribution of Suyeonjang is shown in Table \ref{tab:freq} where two additional music pieces, Songkuyeo and Taryong's node frequency distributions are also shown. 

Imagine that we have a set of $440$ nodes where, for example, the node $n_{18}$ has $76$ copies, $n_{6}$ has $57$ copies and so on. Then the chance of randomly picking up the node $n_j$ can be the same as its probability: 
\[
	\text{ Node probability of $n_j$} = \frac{\text{ Node frequency of $n_j$}} {\text{ Total number of node frequencies}}
\]
In general, consider a music of length $d$ notes that flows in the following order $\mathcal{L} = \{n_1, \ldots, n_d\}$. Let $\{ \nu_1, \ldots, \nu_q\}$ be the set of its distinct nodes as before, with the node frequencies are $f_1, \ldots, f_q$, respectively. 

Then $\mathcal{P}$ for this  music is a multiset made of all the nodes $\nu_1, \ldots, \nu_q$, where node $\nu_1$ appears in the set $f_1$ times, $\nu_2$ appears $f_2$ times and so on.
\[
	 \mathcal{P} = \{ \underbrace{\nu_1, \ldots, \nu_1}_{f_1\text{ times}},  \underbrace{\nu_2, \ldots, \nu_2}_{f_2\text{ times}},  \ldots, \underbrace{\nu_q, \ldots, \nu_q}_{f_q\text{ times}} \}
\]
Notice that 
\[
	f_1 + \ldots + f_q = d,
\]
thus, $\mathcal{P}$ contains exactly $d$ nodes that made up from $q$ distinct nodes from the music. In other words, $\mathcal{P}$ is a permutation of the set  $\mathcal{L} = \{n_1, \ldots, n_d\}$. The chance of picking a node from $\mathcal{P}$ is equal to its probability, $p(\nu_j) = \frac{f_j}{d}$.

\section{Algorithmic composition - Algorithm A}
\label{algorithm}
\begin{figure}[!h]
\begin{center}
\begin{tikzpicture}
\draw (0.5,0) rectangle (4.5,1) node[midway] {Seed music $\mathcal{O}$};
\draw[->,thick] (2.5,0) -- (7,-1);
\draw (5,-1) rectangle (9,-2) node[midway] {Distance matrix $\{\delta_{ij}\}$};
\draw [->,thick] (7,-2) -- (7,-3.5);
\draw (6,-4.5) rectangle (8,-3.5) node[midway] {Cycles $C_i$};
\draw [->,thick] (6,-4) -- (5,-4);
\draw (0,-3.5) rectangle (5,-4.5) node[midway] {Binary Overlap matrix $M_{ij}^s$};
\draw[->,thick] (2.5,0) -- (-1.75,-1);
\draw (-3.5,-1) rectangle (0,-2.5) node[midway,align=center] {Node frequency \\distribution};
\draw [->,thick] (-1.75,-2.5) -- (-1.75,-3.5);
\draw (-3,-3.5) rectangle (-0.5,-4.5) node[midway] {Node pool $\mathcal{P}$};
\draw [->,thick] (-1.75, -4.5) -- (2.3,-6);
\draw [->,thick] (2.5,-4.5) -- (2.5,-6);
\draw [->,thick] (7,-4.5) -- (2.7,-6);
\draw (0.5,-6) rectangle (4.5,-7) node[midway] {New music $\mathcal{O}'$};
\end{tikzpicture}
\end{center}
\caption{Flowchart of algorithmic creation of new music.}
\label{fig:flowchart}
\end{figure}
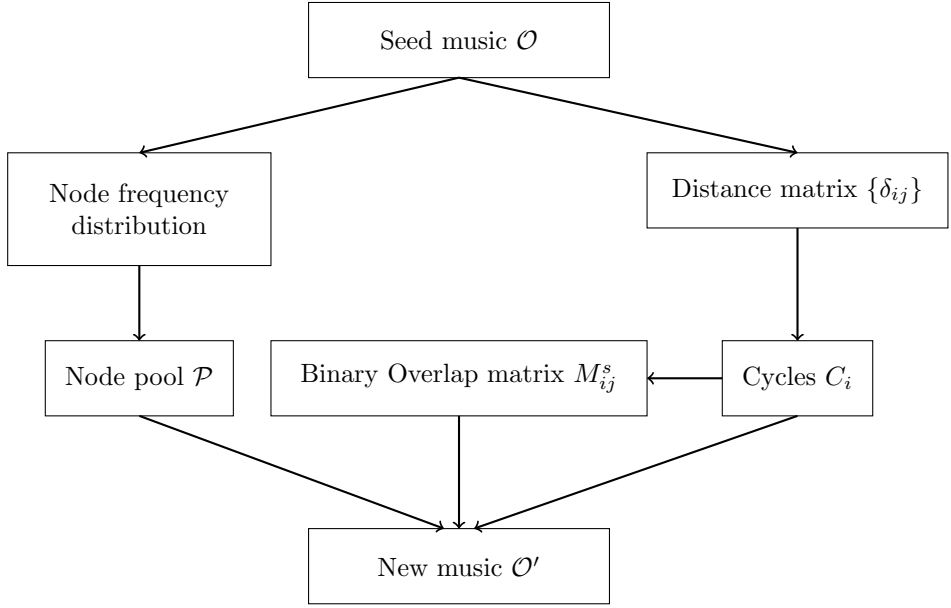

The Overlap matrix of Suyeonjang in  \figref{fig:syj_overlap4} is found to be related to the unique structure of  Dodeuri pattern \cite{TPJ}. The Overlap matrices show how the Cycles are distributed and interconnected along the music flow. The idea of creating new music algorithmically is shown in the flowchart in \figref{fig:flowchart}.

For the preparation we need Node pool, Cycles and binary Overlap matrix. 
Given the music that we consider (Suyeonjang for example), it is straightforward to get the node frequency distribution and then build up the Node pool. On the other hand, from the seed music we can construct the music network and then find the distance matrix. Next, the distance matrix is plugged into Javaplex and by using TDA tools we find the barcode and corresponding Cycle information. The Cycles are then used to obtain the binary Overlap matrix. 

Given the music of length $d$ notes which is a music in Jeongganbo that flows in the following order $\mathcal{L} = \{ n_1, \ldots, n_d\}$, our goal is to algorithmically create new music of the same length $d$ that flows in the following order $\mathcal{L'} = \{ n_1', \ldots, n_d'\}$ such that the pattern of the binary Overlap matrix {\color{black}{of}} given $s$-scale of the seed music is strictly followed and the new music sounds similar to the seed music holding particular patterns. Below we explain how to choose each note $n_j', j = 1,\ldots,d$ of the new music. 

Assume that following the process described above we found $k$ Cycles $C_1, C_2,\ldots, C_k$. Let $S_j$ be the set of Cycles which survive at note $n_j$ {\color{black}{at}} $s$-scale, as defined in \eqref{eq:S}. Along the music flow, at each note $n_j$ either there are some Cycles surviving {\color{black}{at}} $s$-scale ($S_j\neq \emptyset$) or none of the Cycles survive ($S_j=\emptyset$). Denote by $\mathcal{I}_j$ the set of nodes belonging to the intersection of those Cycles surviving {\color{black}{at}} $s$-scale that overlap at note $n_j$:
\[
	\mathcal{I}_j = \{ n_i | n_i \in \bigcap\limits_{C_i \in S_j} C_i \}.
\]

If some of the Cycles survive at note $n_j$ {\color{black}{at}} $s$-scale, i.e., $S_j \neq\emptyset$, then the new note $n_j'$ is randomly chosen from the intersection of those Cycles:
\begin{equation*}
	n_j' = \text{ random choice from } \mathcal{I}_j  \iff S_j \neq \emptyset, j = 1,\ldots,d.
\end{equation*}

Otherwise, if none of the $k$ Cycles survive at note $n_j$ {\color{black}{at}} $s$-scale, i.e., $S_j = \emptyset$, we randomly pick up a node from the Node pool $\mathcal{P}$ with or without a constraint depending on whether or not there exist Cycles surviving at node $n_{j-1}$ and node $n_{j+1}$ {\color{black}{at}} $s$-scale as follows:
\begin{equation*}
\resizebox{ \textwidth}{!} {$
	n_j' = \begin{cases}
	 \text{ random choice from } \mathcal{P}  & \iff S_{j-1} = S_j = S_{j+1} =\emptyset, \\
	 \text{ random choice from } \mathcal{P} \setminus \mathcal{I}_{j-1} & \iff S_{j-1} \neq \emptyset, S_j = S_{j+1} = \emptyset, \\
	 \text{ random choice from } \mathcal{P} \setminus \mathcal{I}_{j+1} & \iff S_{j-1} = S_j = \emptyset, S_{j+1} \neq \emptyset, \\
	 \text{ random choice from } \mathcal{P} \setminus (\mathcal{I}_{j-1} \bigcup \mathcal{I}_{j+1}) & \iff S_{j-1} \neq \emptyset, S_j = \emptyset, S_{j+1} \neq \emptyset,
	\end{cases}$
}
\end{equation*}
for  $j = 1,\ldots,d$. It is easy to see that in this way, we strictly follow the pattern of the binary Overlap matrix.

\begin{remark}
After getting new music $\mathcal{O}'$ we can apply to it the process of constructing music network, followed by using TDA tools, to find its corresponding binary and integer Overlap matrices. It is observed that new music $\mathcal{O}'$ generated by  the procedure in \figref{fig:flowchart}, although sounds nice, neither necessarily has the same number of Cycles nor necessarily reflects the overlap pattern of the original seed music $\mathcal{O}$. In other words, both the binary and integer Overlap matrices of $\mathcal{O}'$ can be very different from those of $\mathcal{O}$. This is illustrated in the following examples where we use Suyeonjang as the seed music. We provide here only the binary Overlap matrices since it is obvious that if two musics have different binary Overlap matrices then their integer Overlap matrices are also different.
\end{remark}
\begin{example} \label{ex:syjnew5}
Figure \ref{fig:syjnew5} shows a music generated from Suyeonjang which has only four Cycles.
\begin{figure}[!h]
	\begin{center}
		\includegraphics[scale=0.8]{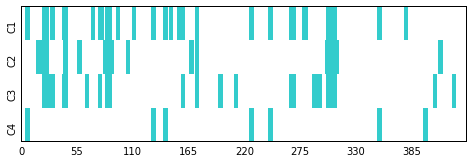}
		\caption{The binary Overlap matrix {\color{black}{of $4$-scale for a new music generated from Suyeonjang}}. The zero entries are left blank and the entries that equal to $1$ are colored. The new music has four Cycles. } 
		\label{fig:syjnew5}
	\end{center}
\end{figure}

\end{example}
\begin{example} \label{ex:syjnew63}
Figure \ref{fig:syjnew63} shows another music generated from Suyeonjang which also has only four Cycles. The overlap pattern of this music  is quite different from that of the music shown in Example \ref{ex:syjnew5}.
\begin{figure}[!h]
	\begin{center}
		\includegraphics[scale=0.8]{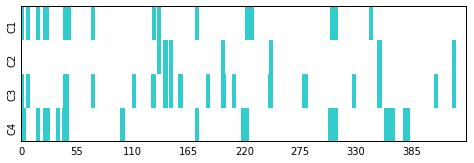}
		\caption{The binary Overlap matrix {\color{black}{of $4$-scale for a new music generated from Suyeonjang}}. The zero entries are left blank and the entries that equal to $1$ are colored. The new music has four Cycles. } 
		\label{fig:syjnew63}
	\end{center}
\end{figure}

\end{example}
\begin{example}
Figure \ref{fig:syjnew68} shows a music generated from Suyeonjang which has six Cycles. Although this music has more Cycles than the musics shown in Examples \ref{ex:syjnew5} and \ref{ex:syjnew63}, it obviously has less Cycles than Suyeonjang. It is not exactly the same as the binary Overlap matrix of Suyeonjang but shows some similarity in the overlapping sense. 
\begin{figure}[!h]
	\begin{center}
		\includegraphics[scale=0.8]{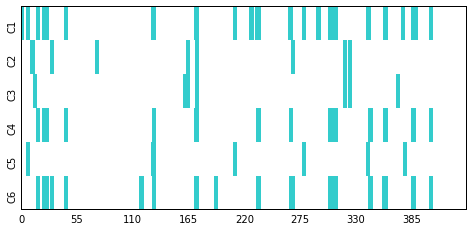}
		\caption{The binary Overlap matrix {\color{black}{of $4$-scale for a new music generated from Suyeonjang}}. The zero entries are left blank and the entries that equal to $1$ are colored. The new music has six Cycles. } 
		\label{fig:syjnew68}
	\end{center}
\end{figure}

\end{example}

\section{Creating new music with artificial neural network - Algorithm B}
\label{ANN}
An alternative approach of generating new music is to use the artificial neural network explained in the following. 

\subsection{Generating seed Overlap matrix}
\label{seed}
Given  the integer Overlap matrix {\color{black}{of $s$-scale for Suyeonjang}} $M_{k\times d}^s$, 
our first goal is to generate an integer Overlap matrix $\widetilde{M}_{k\times d}^s$ that has the same size and similar pattern as $M_{k\times d}^s$, which will be used as a seed Overlap matrix towards the artificial neural network. 

For the given music, there could be various ways of generating seed Overlap matrix which has the same size as the Overlap matrix of the given music and also has similar patterns as the given music. Below we introduce three algorithms for generating a seed integer Overlap matrix $\widetilde{M}_{k\times d}^s$ from $M_{k\times d}^s$.

{\color{black}{
The common strategy of the following three algorithms is to generate a binary Overlap matrix that has the same size and mimic the overlapping pattern of the given integer Overlap matrix first, then convert it to an integer Overlap matrix. The idea of each algorithm is as follows. 

{\em Row by Row Method:} Overlap Matrix Algorithm \#1 is row by row approach. That is, the first binary row is determined based on the number of blocks of consecutive nonzero entries of the first row of the given Overlap matrix. Next, using the overlapping pattern and node information, the second binary row is generated so that it overlaps or does not overlap the first row depending on whether the first and the second rows of the given Overlap matrix overlap or not. This process is continued for all rows. As a result we get a binary Overlap matrix that has the same size, same frequency of blocks of consecutive nonzero entries and preserve the overlapping pattern of the given Overlap matrix. To convert the generated binary Overlap matrix to an integer Overlap matrix, we convert column by column using the node information.

{\em Element by Element {\color{black}{Method}}:} Overlap Matrix Algorithm \#2 is element by element approach. That is, nonzero entries in given Overlap matrix are first replaced with 1 to generate a binary Overlap matrix that has exactly the same overlapping pattern as the given Overlap matrix. Then, from the node information, entries equal to 1 in the generated binary Overlap matrix are converted back to integer numbers. This algorithm is the simplest one if we just want to get a new but very similar Overlap matrix.

{\em Column by Column Method:} Overlap Matrix Algorithm \#3 is column by column approach. After converting given integer Overlap matrix to a binary Overlap matrix, we collect all kinds of columns in it and generate new Overlap matrix according to the frequencies of the columns. {\color{black}{In this}} way we automatically preserve the overlapping pattern of the given Overlap matrix, while still have the flexibility in the number of blocks as well as the length of each block of consecutive nonzero entries. The difficulty in this algorithm is that, a new column has to be carefully chosen so that the number of consecutive nonzero entries is not less than $s$ to satisfy the definition of an Overlap matrix {\color{black}{of}} $s$-scale.  
}}

\vspace{0.4cm}
\noindent
\rule{\textwidth}{1pt}
\vspace{-0.8cm}
\begin{center}
	Overlap Matrix Algorithm \#1 
\end{center}	
\vspace{-0.1cm}
\hrule
\vspace{0.3cm}
\vspace{0.5cm}

\begin{enumerate}
	\item [Step 1:] Find the frequency $f_i$ of blocks of consecutive nonzero entries in row $i$ of $M_{k\times d}^s$, $i=1,\ldots,k$.
	\item [Step 2:] For each cycle $C_i$, identify the set of cycles that do not overlap $C_i$. Let $S_i$ be the set of all indices $j$ of nodes $n_j$ that constitute those cycles.
	\item [Step 3:] Let $\widetilde{B}_{k\times d}^s$ be a $k\times d$ zero matrix. Then for each $i$ randomly pick up $f_i$ indices $j_1,\ldots, j_{f_i}$ that is not in $S_i$ and set $\widetilde{b}_{i,j_p+q} = 1$, $p = 1,\ldots, f_i$;  $q = 0,1,\ldots,s-1$. Repeat this for all $i = 1,\ldots,k$.  This step generates $\widetilde{B}_{k\times d}^s$ as a binary {\color{black}{Overlap}} matrix.
	\item [Step 4:] For each non-zero column $\bold{b_j}$ of $\widetilde{B}_{k\times d}^s$ let $U_j$ be the set of all nodes that constitute those cycles that correspond to nonzero entries in $\bold{b_j}$. Then replace all entries equal to $1$ in $\bold{b_j}$ by a random node index $n_j$ chosen from $U_j$. This step converts $\widetilde{B}_{k\times d}^s$ to an integer {\color{black}{Overlap}} matrix $\widetilde{M}_{k\times d}^s$.
\end{enumerate}

\vspace{1.6cm}
\hrule


\noindent
\rule{\textwidth}{1pt}
\vspace{-0.8cm}
\begin{center}
	Overlap Matrix Algorithm \#2 
\end{center}	
\vspace{-0.1cm}
\hrule
\vspace{0.3cm}
\vspace{0.5cm}

\begin{enumerate}
	\item [Step 1:] Replace nonzero entries in $M_{k\times d}^s$ with $1$ to convert $M_{k\times d}^s$ to a binary Overlap matrix $B_{k\times d}^s$.
	\item [Step 2:] Assume that $C_i = \{ n_1^i, n_2^i,\ldots,n_{r_i}^i\}, i = 1,\ldots,k$. Replace entries equal to $1$ in row $i$ of $B_{k\times d}^s$ by a random node $n_j^i$ belonging to $C_i$ by the method \texttt{random.choice(C\_i)}. 
	\item [Step 3:] Repeat Step 2 for all rows in $B_{k\times d}^s$.
\end{enumerate}

\vspace{0.3cm}
\hrule

\noindent
\rule{\textwidth}{1pt}
\vspace{-0.8cm}
\begin{center}
	Overlap Matrix Algorithm \#3 
\end{center}	
\vspace{-0.1cm}
\hrule
\vspace{0.3cm}
\vspace{0.5cm}

\noindent\textbf{Procedure 1}: Generate a binary Overlap matrix $\widetilde{B}_{k\times d}^s$.
\begin{enumerate}
	\item [Step 1:] Convert $M_{k\times d}^s$ to a binary Overlap matrix $B_{k\times d}^s$.
	\item [Step 2:] Choose a column for $\widetilde{B}_{k\times d}^s$ by a random selection from the set of all distinct columns in $B_{k\times d}^s$ according to the frequencies of the columns by the method \texttt{random.choices(C,F)}, where $C=[c_1,\ldots,c_i]$ denotes the list of all distinct columns in $B_{k\times d}^s$ and $F = [f_1,\ldots,f_i]$ is the list of their corresponding frequencies.
	\item [Step 3:] Select a number for how many times the column selected in Step 2 will be repeated in $\widetilde{B}_{k\times d}^s$. This number is chosen by \texttt{random.choices} from the set  of repeated columns according to the repeating frequencies.
	\item [Step 4:] The next column is also chosen by the method \texttt{random.choices} but from the set of all distinct columns  in $B_{k\times d}^s$ that are adjacent to the column selected in Step 2. The resulting column in Step 4 after placing in $\widetilde{B}_{k\times d}^s$ should satisfy the continuous condition, meaning that any entry equal to $1$ in $\widetilde{B}_{k\times d}^s$  should be staying in a consecutive sequence of length at least $s$ columns that equal to $1$.
	\item [Step 5:] Repeat Step 2-4 if necessary until the frequencies of blocks of consecutive entries equal to $1$ on each row of $B_{k\times d}^s$  and  $\widetilde{B}_{k\times d}^s$ are similar. 
\end{enumerate}
\noindent \textbf{Procedure 2}: Convert $\widetilde{B}_{k\times d}^s$ to an integer Overlap matrix $\widetilde{M}_{k\times d}^s$.
\begin{enumerate}
\item [Step 1:] Find all kinds of columns that $M_{k\times d}^s$ has and save them in a variable named \texttt{col\_choice}.
\item [Step 2:] For each binary column $\bold{b_j}$ in $\widetilde{B}_{k\times d}^s$, find all columns in \texttt{col\_choice} that have the same indices of nonzero entries as $\bold{b_j}$, then randomly choose one of them, say $\bold{\widetilde{m_j}}$, to add to $\widetilde{M}_{k\times d}^s$.

\end{enumerate}

\vspace{0.3cm}
\hrule
\subsection{Training data set}
In order to input the seed Overlap matrix into the artificial neural network for generating a new music, we need to construct an optimized artificial neural network corresponding to the given music. Such network can be obtained by training a machine with the given music. Note that the number of the given music is not large. For example, if we consider a single music, e.g. Suyeonjang, the number of the input music with which we train the machine is simply unity. That is, the number of data required for training is too small. In this section, we use a simple method of the {\it periodic extension} of the given music to generate more music for training. In our future research, however, we need to investigate more general approaches to address the issue of the number of the input music. The periodic extension is reasonable approach reflecting the Dodeuri music patterns. 

\begin{figure}[!h]
    \centering
    \includegraphics[width=0.8\textwidth]{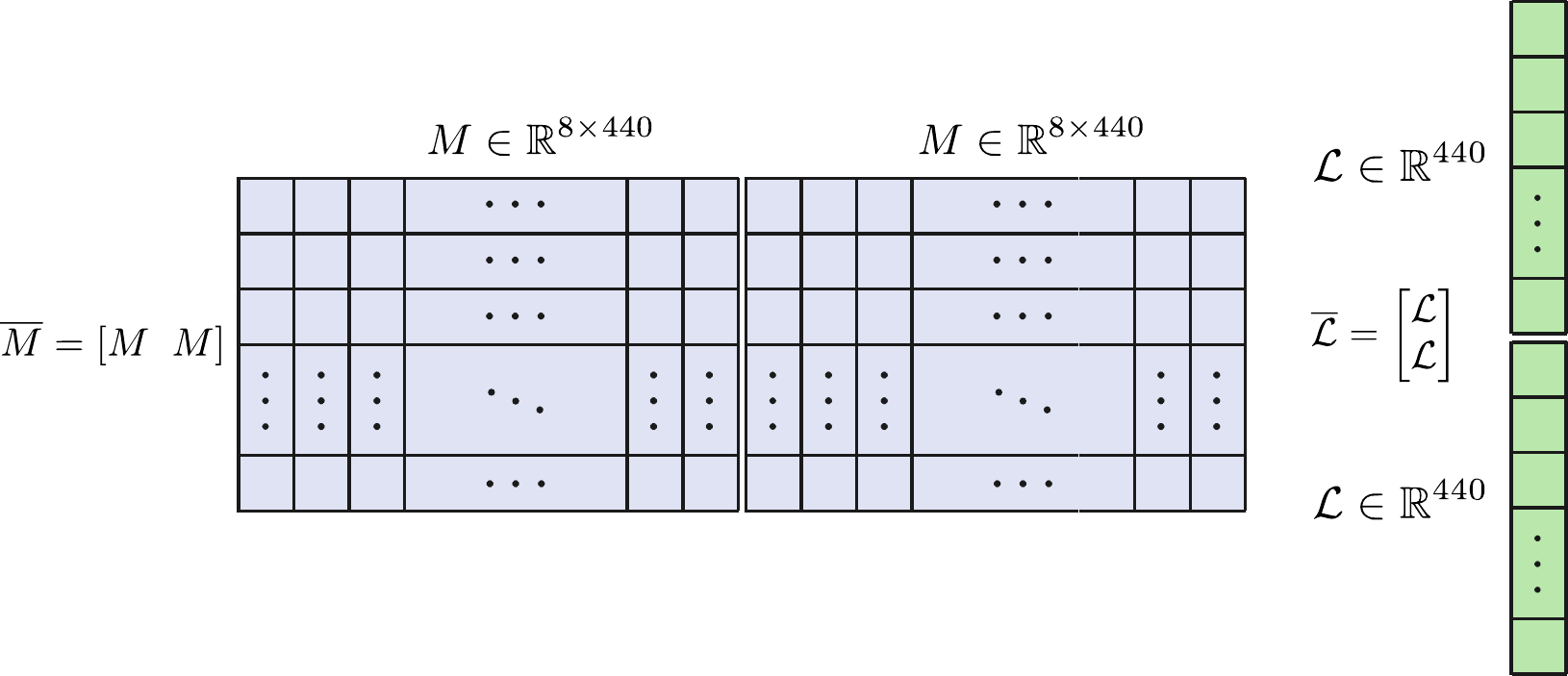}
    \caption{Dataset construction: periodic extension}
    \label{fig:dataset}
\end{figure}

Let $M_{k\times d}^s$, which from now on will be denoted as $M$, be the integer Overlap matrix of the given music {\color{black}{and $\mathcal{L}$ be the ordered sequence of notes as the music flow.}} 
We augment seed music by shifting the original music one space at a time:
\begin{equation}
\begin{matrix}
M^{(i)} & = & \overline{M}_{:,i:i + d - 1}, \\
\mathcal{L}^{(i)} & = & \overline{\mathcal{L}}_{i:i + d - 1},
\end{matrix}
\label{eq:augmentation}
\end{equation}
where 
$\overline{M} = [{\color{black}{M \;\; M}}]$, 
$\overline{\mathcal{L}} = \begin{bmatrix}\mathcal{L} \\ \mathcal{L} \end{bmatrix}$, $d$ is the length of a seed music {\color{black}{and $i = 1,\ldots,d$}}. 
For the illustration, let us use Suyeonjang as the seed music. As explained above, for this case, we have $d = 440$ and $k = 8$ for Suyeonjang. 
Here, we think of a music flow $\mathcal{L}$ as a vector whose elements are indices of notes. Let $\mathcal{X}=\left\{ \overline{M}_{:,i:i+439}\right\}_{i=1}^{440}$ and $\mathcal{Y}=\left\{ \overline{\mathcal{L}}_{i:i+439}\right\}_{i=1}^{440}$.  Figure \ref{fig:dataset} illustrates the construction of our dataset.

We design a neural network $f_\theta$ such that $f_\theta(M') = {y}'$ where $M' \in \mathcal{X}$ and $y' \in \mathcal{Y}$ . Note that $f_\theta$ only satisfies $f_\theta(M') = {y}'$ in the optimization sense. 


\subsection{Construction of music generation network $f_\theta$}
%
For the construction of $f_\theta$ introduced above, we seek a set of parameters $\theta^{*}$ that maximizes the probability of the real music flow $\mathcal{L}$ given the integer Overlap matrix $M^{s}_{k \times d}$:
$$
    \theta^{*} = \operatornamewithlimits{argmax}_{\theta} \sum_{(\mathcal{L}^{(i)}, M^{(i)})}\log p\left( \mathcal{L}^{(i)} \mid M^{(i)} \right),
$$
where $\theta$ is a set of parameters of our model, {\color{black}{$p$ is the conditional probability distribution and}} $(\mathcal{L}^{(i)}, M^{(i)})$ is the $i$-th pair of the real music flow and its corresponding integer Overlap matrix induced by $\mathcal{L}$ and $M^{s}_{k \times d}$.  We model the conditional probability distribution $p$ with a Multi-Layer Perceptron (MLP), but one can use any nonlinear function. MLPs are a sequence of affine transformations followed by element-wise nonlinearity. Let $f^{(l)}$ be the $l$-th hidden layer of a MLP and let $\mathbf{a}^{(l-1)} \in \mathbb{R}^{d^{(l-1)}}$ be an input vector of $f^{(l)}$ where $d^{(l-1)}$ is the dimensionality of $(l-1)$-th hidden layer. The output of $f^{(l)}$ is:
$$
f^{(l)} \left( \mathbf{a}^{(l-1)} \right) = \sigma \left( \mathbf{W}^{(l)} \mathbf{a}^{(l-1)} + \mathbf{b}^{(l)} \right),
$$
where {\color{black}{$\sigma$ is an activation function}}, $\mathbf{W}^{(l)} \in \mathbb{R}^{d^{(l)} \times d^{(l-1)}}$ and $\mathbf{b}^{(l)} \in \mathbb{R}^{d^{(l)}}$ are learnable weight matrix and bias vector, respectively. A nonlinear function $\sigma$ is applied to each element of a vector. Each hidden layer can use a different activation function. Typical choices for $\sigma$ are sigmoid function, hyperbolic tangent function and Rectified Linear Unit (ReLU). Then, a MLP $f_\theta$ is the composition of hidden layers $f^{(l)}$s:
$$
\begin{matrix}
  f_\theta & = & f^{(L)} \circ f^{(L-1)} \circ \cdots \circ f^{(1)}, \\
  \theta & = & \left\{ \mathbf{W}^{(l)}, \mathbf{b}^{(l)} :l=1, \cdots, L \right\},
\end{matrix}
$$
where $L$ {\color{black}{is}} the number of hidden layers and $\mathbf{a}^{(0)}=\mathbf{x}$ is an input vector of $f_\theta$. For MLPs, the number of hidden layers, $L$ and the dimensionality of each hidden layer, $d^{(l)}$ are hyper-parameters to be determined. 


%

\begin{figure}[htbp]
    \centering
    \includegraphics[width=\textwidth]{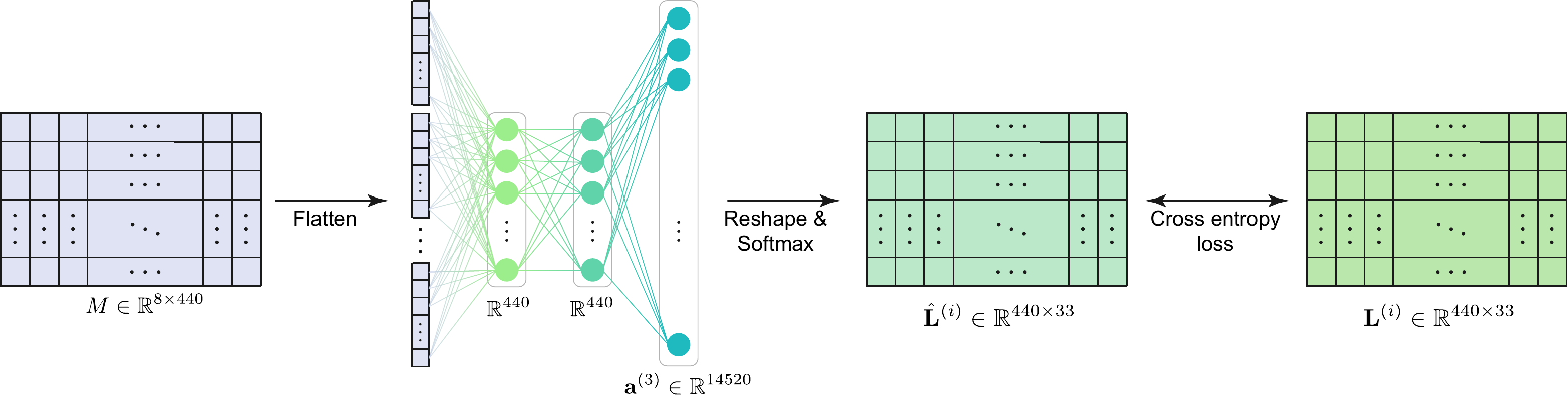}
    \caption{The architecture of our model for Suyeonjang. We use a MLP with three hidden layers each of which has 440 dimensionality. The output is the 440 probability distributions over 33 distinct notes. The learnable parameters are updated toward minimizing the cross entropy loss.}
    \label{fig:network}
\end{figure}

In general, MLPs take a vector as an input while our input is a matrix. The simplest way to feed a matrix into a MLP is to flatten it to one dimensional vector. Our model $f_\theta$ takes the flattened vector of $M^{(i)}$ and outputs the $d$ probability distributions \textcolor{black}{over $q$ distinct notes. We note that each element of $\mathcal{L}^{(i)}$ is the node index so that we apply one-hot encoding to it. Hence, we generate the corresponding $d \times q$ matrix $\mathbf{L}^{(i)}$ such that $\mathbf{L}^{(i)}_{jk}=1$ if $j$-th note is equal to $\nu_k$, and $\mathbf{L}^{(i)}_{jk}=0$ otherwise. For our network, the output of the last hidden layer $\mathbf{a}^{(L)}$ is a $dq$-dimensional vector. Then, we reshape $\mathbf{a}^{(L)}$ into the $d \times q$ matrix $\hat{\mathbf{L}}^{(i)}$, and takes the softmax function over each row. The output of our model for $M^{(i)}$ is as follows:}
\begin{equation}
\hat{\mathbf{L}}^{(i)}_{jk}=\frac{\exp \left( \mathbf{a}^{(L)}_{q(j-1) + k} \right)}{\sum_{l=1}^{q} \exp \left( \mathbf{a}^{(L)}_{q(j-1) + l} \right)}.
\label{eq:softmax}
\end{equation}
$\hat{\mathbf{L}}^{(i)}_{jk}$ can be interpreted as the probability that the $j$-th note in the generated music is $\nu_k$. Then, a set of parameters of our model is updated toward minimizing the cross entropy loss between output probability distributions and the real music flow:
\begin{equation}
\operatorname{CrossEntropyLoss}\left(\hat{\mathbf{L}}^{(i)}, \mathbf{L}^{(i)} \right) = - \frac{1}{d} \sum\limits_{j=1}^{d} \sum\limits_{k=1}^{q} \mathbf{L}^{(i)}_{jk} \log \hat{\mathbf{L}}^{(i)}_{jk}.
\label{eq:cross-entropy}
\end{equation}
Figure \ref{fig:network} shows the architecture of our model.

To evaluate our model, {\color{black}{we}} generated Suyeonjang-style musics using our model. The length $d$ of Suyeonjang is 440 and it has $q=33$ distinct notes and $k=8$ cycles. We used the binary Overlap matrix {\color{black}{of}} 4-scale. We obtained 440 data from the augmentation in Equation {\color{black}{\eqref{eq:augmentation}}} and used $70\%$ of them for training and the rest of them for evaluation. The detailed architecture of a MLP is shown in Table \ref{tab:architecture}. We optimized Equation {\color{black}{\eqref{eq:cross-entropy}}} with respect to the model's parameters using Adam optimizer with learning rate $0.001$ over 500 epochs.  

\begin{table}[htbp]
    \centering
    \begin{tabular}{c|c|c}
         \hline
         $l$ & $d^{(l)}$ & $\sigma$ \\
         \hline
         $1$ & $440$ & ReLU \\
         $2$ & $440$ & ReLU \\
         $3$ & $440 \times 33$ & Softmax in Eq. {\color{black}{\eqref{eq:softmax} }}\\
         \hline
    \end{tabular}
    \caption{The architecture of the MLP for generating Suyeonjang-style music}
    \label{tab:architecture}
\end{table}

\subsection{Examples}
\begin{figure}[htbp]
    \centering
    \includegraphics[width=\textwidth]{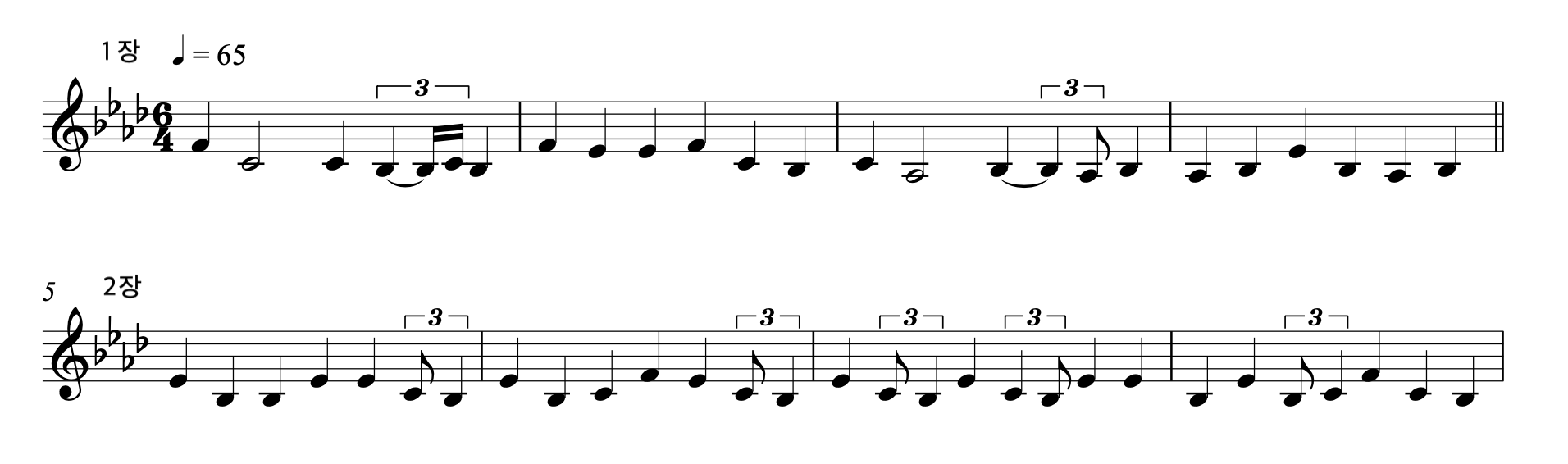}
     \includegraphics[width=\textwidth]{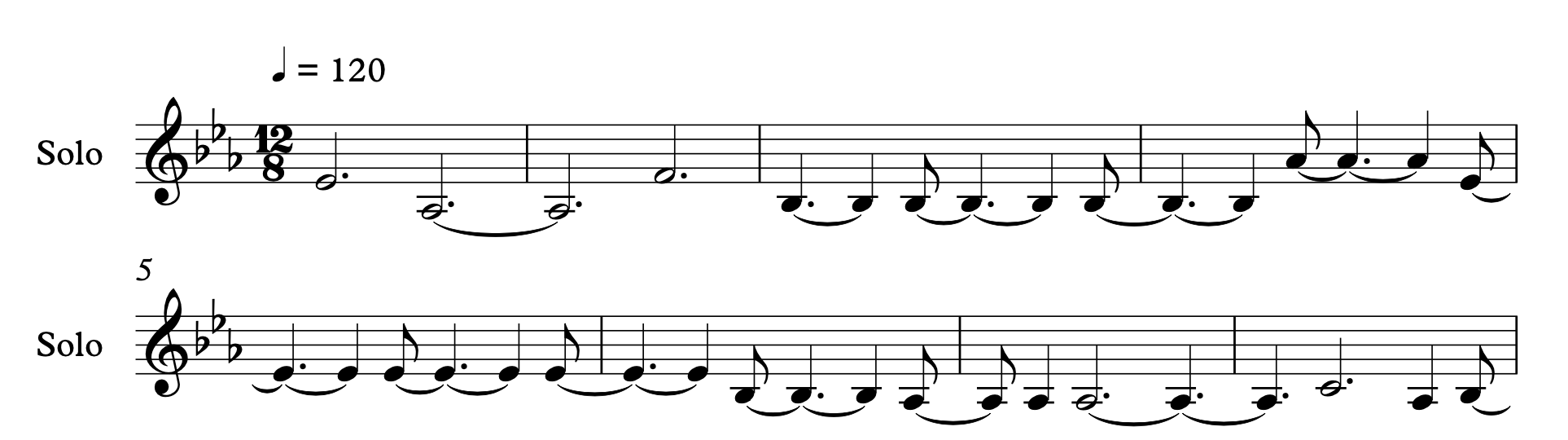}
      \includegraphics[width=\textwidth]{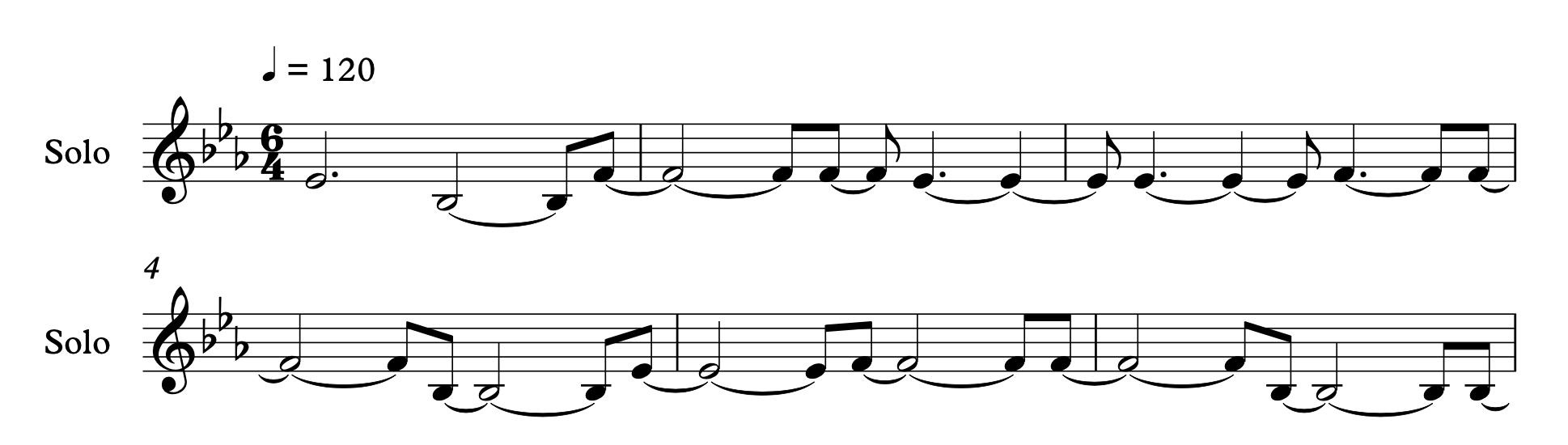}
    \caption{Top: First few notes from the original Suyeonjang with Haegeum instrument. Middle: A music created by Algorithm A. Bottom: A music created by Algorithm B.}
    \label{fig:example}
\end{figure}

We generated music pieces with Algorithm A and Algorithm B.\footnote{
Some of the generated music pieces were played in June and July, 2021. Readers can listen those using the following YouTube links: {\lstinline{https://www.youtube.com/watch?v=_DKo8FjL7Mg&t=461s}} (June 5, 2021) and {\lstinline{https://www.youtube.com/watch?v=AxXKoFRlQiQ&t=751s}} (July 29, 2021). The original music with Haegeum instrument is played from 0:24 to 5:24 in the first link and 0:10 to 4:52 in the second link. } 
For Algorithm B, we used the Overlap Matrix Algorithm \# 1. 
Figure \ref{fig:example} shows the original Suyeonjang for Haegeum instrument (top), one of the generated music pieces with Algorithm {\color{black}{A}} (middle), and one of the generated music pieces with Algorithm {\color{black}{B}} (bottom). The music pieces in the middle and bottom are randomly selected from the automatically generated music pool. 

\section{Conclusion}
\label{conclusion}
In this paper, we used topological data analysis, {\color{black}{Overlap}} matrix and artificial neural network approaches for machine composition of trained Korean music, particularly the {\it {Dodeuri}} music. Using the concept of training the composition principle, we could generate similar music pieces to Dodeuri music. Although the proposed method provides a framework of  machine composition of Korean music, there are several issues that need further rigorous investigations. First of all, we will need to analyze the overall structures of the generated music through Algorithm A and Algorithm B, compare them with the original music and study its musical implications. Second, the current research considered only limited aspects of Korean music reflected on the Overlap matrix, but for a full consideration, we will need to consider other unique characteristics of Korean music such as meter, ornamenting symbols, {\it Sikimse}\footnote{Sikimse is a unique technique of Korean music that variates the given note by vibrating, sliding, breaking, etc. }, etc. Also, the current research used a rather simple periodic extension method for generating the training data set from the given seed music.  Our future work will conduct a study on how to provide training data when the number of considered music pieces is small. These should be fully considered for the construction of more generalized machine composition of Korean music. 

\vskip .2in
\noindent
{\bf{Acknowledgements:} }This work was supported by the NRF grant under grant number  2021R1A2C300964812. We thank Dr. Myong-Ok Kim for her translating Suyeonjangjigok into  staff notation (\figref{fig:syj_original} and top figure of  \figref{fig:example}). 

\newpage

\bibliographystyle{plain}
\bibliography{comj_ref}




\end{document}